\documentclass[aps,prd,twocolumn,showpacs,amsmath,amssymb,nofootinbib]{revtex4-2}
\usepackage{graphicx}
\usepackage{color}
\usepackage{times}
\usepackage{inputenc}
\usepackage{bm}
\usepackage{ulem}
\usepackage{multirow}
\usepackage{float}
\usepackage{url}
\usepackage{natbib}
\usepackage{mathrsfs}
\usepackage{physics}
\usepackage{comment}
\usepackage[table,xcdraw]{xcolor}
\usepackage{booktabs,makecell}
\usepackage{comment}
\usepackage[colorlinks=true,citecolor=,urlcolor=blue,linkcolor=blue]{hyperref}
\usepackage[caption=false]{subfig}
\usepackage{booktabs}
\usepackage{multirow}

\newcommand{\apjl}{Astro. Phys. J. Lett.}
\newcommand{\mnras}{Mon. Not. Roy. Astron. Soc.}

\newcommand{\nphysa}{Nuc. Phys. A}

\newcommand{\sovast}{Soviet Astronomy}

\begin{document}

%\preprint{APS/123-QED}

\title{Bayesian Inference of dense matter equation of state of  neutron star with antikaon condensation}% Force line breaks with \\

\author{Vishal Parmar$^{1}$}
\email{vishalparmar@iitj.ac.in}
\author{Vivek Baruah Thapa$^{2}$}
\email{Corresponding author: vivek.thapa@bacollege.ac.in}
\author{Anil Kumar$^{1}$}
\author{Debades Bandyopadhyay $^{3}$}
\author{Monika Sinha$^{1}$}
\affiliation{\it $^{1}$ Indian Institute of Technology Jodhpur, Jodhpur 342037 India}
\affiliation{\it $^{2}$ Department of Physics, Bhawanipur Anchalik College, Barpeta, Assam 781352, India}
\affiliation{\it $^{3}$ Department of Physics, Aliah University, New Town - 700160, India}

\date{\today}% It is always \today, today,

\begin{abstract}
In this paper, we employ the Density Dependent
Relativistic Hadron (DDRH) field theoretical  Model in a Bayesian analysis to investigate the equation of state (EOS) of dense matter featuring antikaon condensation for $K^-$ and $\bar{K}^0$ inside neutron stars. The vector coupling parameters within the kaonic sector are determined through the iso-spin counting rule and quark model. Our study integrates various constraints, including $\chi$EFT calculations, nuclear saturation properties, and astrophysical observations from pulsars PSR J0030+0451 and PSR J0740+66 and from the GW170817 event. We present posterior distributions of model parameters derived from these constraints, enabling us to explore the distributions of nuclear matter properties and neutron star (NS) characteristics such as radii, tidal deformabilities, central energy densities, and speed of sound. The antikaon potential at the 68(90)\% confidence intervals is determined to be $-129.36^{+12.53(+32.617)}_{-3.837(-5.696)}$ MeV.
\textcolor{black}{This aligns with several studies providing estimates within the range of $-120$ to $-150$ MeV.}
We find that the maximum neutron star mass is constrained to around 2M$_\odot$ due to the significant softening of the EOS caused by antikaon condensation. This softening results in a considerable decrease in the speed of sound. Although antikaon condensation for $K^-$ is not feasible inside the canonical neutron stars, it becomes feasible for higher NS masses. The condensation of both $K^-$ and $\bar{K}^0$ is probably present in the interior of neutron star with mass greater than 2M$_\odot$. We also discuss the interconnections among input variables, isoscalar and isovector aspects of the EOS, and specific NS properties in the context of antikaon condensation. 
\end{abstract}

\maketitle

%\tableofcontents

\section{Introduction}
The era of multimessenger astronomy coupled with ongoing advancements in nuclear theory and instrumentation, including the emergence of third-generation gravitational wave observatories like the Einstein Telescope (ET) and Cosmic Explorer (CE), has unveiled a fresh perspective into the profound fundamental physics of NSs. NSs are remnants of core-collapse supernovae, typically with masses $1.17 - 2$ times that of the sun, confined within a $11-13$ km radius \cite{Capano_2020}. Understanding these extreme celestial objects requires a multidisciplinary approach involving nuclear physics, astrophysics, relativists, condensed matter physics, and experimentalists. The NS structure, comprising the crystalline solid outer and inner crust and the liquid homogeneous core, poses tremendous modelling challenges due to the vast density variations (10-15 order) across these layers \cite{Chamel2008, PhysRevLett.102.191102, Lattimer_2004}. Theoretical models of dense matter in a steller environment, which are essential in describing NSs' interior, must align with terrestrial nuclear experiments and astrophysical data, such as mass-radius measurements, surface temperature, spin, and magnetic field evolution, among others. By correlating these multifaceted observations with theoretical predictions, we can refine and validate our understanding of NSs' fundamental properties and associated EOS. There have been countless efforts in the last few decades to devise an EoS based on theoretical models governing NSs, which still remains a task to solve \cite{Huth2022, Lattimer_2021}.

The nuclear matter composition within a neutron star is predominantly neutrons, with a small proportion of protons and electrons at lower densities, transitioning to electrons alongside muons at higher densities, all in a state of $\beta$-equilibrium and charge neutrality. However, within the star's interior where density reaches extreme levels, new hadronic degrees of freedom like hyperons \cite{Ambartsumyan_1960, apo_2010, Vidana_2000} and $\Delta$-resonances \cite{Cai_2015, LI2018234}
are expected to emerge alongside nucleons. Researchers have been consistently studying these phenomena over the last few decades to understand their mechanisms and assess their occurrence feasibility, aligning with both nuclear matter data and astrophysical observations \cite{Sedrakian_2020, Dexheimer_2021, LI2020135812}. Additionally, beyond these mentioned degrees of freedom, another possibility at high densities is the emergence of various meson condensates as a new non-nucleonic degree of freedom \cite{Haensel_1982}. \textcolor{black}{At extremely high densities, new particles may emerge, and/or matter may transform into a new phase characterized by approximate chiral symmetry restoration, commonly referred to as the hadron-quark phase transition \cite{Schaffner-Bielich_2020, Takatsy_2023, Annala2020, Annala2023}. Currently the hadron-quark phase transition and other possibilities are beyond the scope of present work, but ongoing efforts are being made in this direction}

As density increases, if the electron Fermi energy aligns with the in-medium energy of mesons such as pions or kaons, a condensate of mesons can form. While the occurrence of $\pi$ mesons is improbable \cite{Glendenning_1985, Bandyopadhyay:2021gnp}, the attractive interaction of anti-K mesons with nucleons reduces their effective ground state mass and energy, making the appearance of K mesons possible. Consequently, antikaons may undergo a s-wave (p=0) Bose-Einstein condensation in dense matter created during heavy-ion collisions \cite{Kaplan:1987sc, Nelson:1987dg}. Various authors have examined the signature attractive of antikaon-nucleon interaction in K$^-$ atomic data and antikaon-nucleon scattering data, using chiral perturbation theory, which supports the idea of a s-wave K$^-$ condensed phase existing within neutron stars \cite{Lutz:2000us, 1994PhLB..326...14L}. On the other hand, K$^+$ condensation is not favoured within neutron stars due to its repulsive optical potential nature in nuclear matter \cite{Vivek_2020, Thapa_2021}.

The aim of this study is to develop a comprehensive model framework based on a phenomenological approach using density-dependent couplings with mesons to describe nuclear matter properties, particularly focusing on the emergence and impact of antikaon condensation as a new degree of freedom within neutron star and quantifying the most probable antikaon potential. Drawing insights from existing models and constraints, including observational data and theoretical frameworks like chiral effective field theory ($\chi$EFT), we aim to understand how the introduction of antikaon condensation affects the nuclear matter EOS and stellar properties, such as mass-radius configurations, tidal deformability, and speed of sound. Building upon the success of the Density Dependent Relativistic Hadron (DDRH) field theoretical approach in 
%capturing quantal fluctuations (Is this for meson  fields?) and
describing nuclear matter properties accurately \cite{DD2, DDME2, Hofmann_2001}, we will extend this framework to incorporate kaon condensation phenomena. Through Bayesian statistical analysis and model parameterization, we will explore the density-dependent couplings of vector, isovector and isoscalar fields, akin to the Lorentz-invariant Lagrangian formulations seen in previous studies.

In the past, several attempts have been made to extract the antikaon-nucleus potential, which includes analysis of kaonic-atom data \cite{FRIEDMAN1994518}, enhanced $K^-/K^+$ ratio measured in Heavy-ion collisions \cite{PhysRevLett.91.152301}, self-consistent calculations based on a chiral lagrangian \cite{SCHAFFNERBIELICH2000153, Mishra2009} or meson-exchange potentials \cite{TOLOS2001547}, differential data on elastic $K^-A$ scattering \cite{sibirtsev1999probing}, etc. While the best-fit analysis of kaonic-atom data suggests very strong attractive well depths of around 200 MeV at normal nuclear matter density, self-consistent calculations using a chiral Lagrangian or meson-exchange potentials predict more moderate attractive depths of 50 to 80 MeV. Additionally, studies of kaonic atoms employing the chiral $\bar{K}N$ amplitudes show that the data can be reasonably reproduced with a relatively shallow antikaon-nucleus potential \cite{OSET199899}, with the inclusion of a moderate phenomenological adjustment  \cite{BACA2000335}. These studies also indicate that kaonic atom data do not adequately constrain the antikaon-nucleus potential at normal nuclear matter density \cite{PhysRevC.65.054907, CIEPLY2001173, Magas2011}. Heavy-ion collisions also offer insights into the modification of hadron properties. Specifically, the enhanced $K^-/K^+$ ratio  can be interpreted as evidence of strong antikaon attraction, though alternative explanations involving enhanced cross sections due to the $\Lambda$(1405) resonance shifting to higher energies have been proposed \cite{PhysRevC.65.054907}. In the literature, the antikaon nuclear potential has also been explored in the contexts of hot and dense matter \cite{PhysRevC.65.054907}, asymmetric hyperonic matter \cite{Mishra2009}, and neutron stars \cite{Thapa_2021} to better understand its influence. Given the wide range of antikaon potential values adopted or measured in experimental and theoretical calculations, there is a need to constrain it more precisely. Since kaonic atom data are insufficient for this purpose, researchers can turn to multimessenger observations of neutron stars, leveraging advancements in astronomical instrumentation and analysis techniques.

Our study will involve generating a set of models constrained by $\chi$EFT calculations, nuclear saturation properties, and the astrophysical constraints by PSR J0030+0451 and PSR J0740+66 and GW170817 event. By inferring model parameters and conducting detailed statistical analyses, we aim to investigate how the inclusion of kaon condensation impacts NS properties compared to nucleonic hypotheses. This includes evaluating maximum mass configurations, radius variations, tidal deformability, and central baryonic densities under different scenarios, including those with nucleons alone and those incorporating kaon condensation as a new degree of freedom. Furthermore, we will compare our results with existing observations from instruments like NICER and LIGO/Virgo, exploring compatibility with different scenarios of kaon condensation and their effects on NS properties. The study aims to contribute insights into the role of antikaon condensation in NS matter, shedding light on its implications for stellar structure and EOS under extreme conditions.

The paper is organized as follows. In Section \ref{formalism}, we briefly discuss the DDRH field theory formalism incorporated in this work. We introduce the priors and constraints for our Bayesian analysis. The results are presented in Section \ref{results}, where we discuss the posterior distribution of various nuclear matter and neutron star properties and associated correlations. We summarize our work in Section \ref{summary}.
%\clearpage

\section{\label{formalism} Formalism}
In this section, we will begin by detailing the formalism of the EOS utilizing the DDRH model. Subsequently, we will discuss the Bayesian framework, discussing the nuclear matter and astrophysical constraints that have been incorporated into the current study.

\subsection{CDF Model for dense matter}

In this segment, we present the density-dependent model for analyzing the phase shift from hadronic to antikaon condensed material, which may manifest in either first-order or second-order configurations. Throughout our investigation, we've included nucleons ($N\equiv n,p$) in addition to electrons and muons within the hadronic substance. The strong interactions among the baryons and antikaons are facilitated through the scalar $\sigma$, isoscalar-vector $\omega^\mu$, and isovector-vector $\rho^{\mu \nu}$ meson fields.
We have considered the mean field model with  density-dependent coupling constants. Throughout the model, the implementation of natural units is incorporated ($\hbar=c=1$). In general, the total Lagrangian density of the matter is given by  \cite{1999PhRvC..60b5803G,2000NuPhA.674..553P,2001PhRvC..64e5805B, Sarmistha_2001, 2000NuPhA.674..553P,1999PhRvC..60b5803G}
\begin{equation}\label{rmftlagrangian}
\begin{aligned}
\mathcal{L} & = \mathcal{L}_N + \mathcal{L}_K + \mathcal{L}_l + \frac{1}{2}(\partial_{\mu}\sigma\partial^{\mu}\sigma - m_{\sigma}^2 \sigma^2) - \frac{1}{4}\omega_{\mu\nu}\omega^{\mu\nu} \\
&  + \frac{1}{2}m_{\omega}^2\omega_{\mu}\omega^{\mu} - \frac{1}{4}\boldsymbol{\rho}_{\mu\nu} \cdot \boldsymbol{\rho}^{\mu\nu} + \frac{1}{2}m_{\rho}^2\boldsymbol{\rho}_{\mu} \cdot \boldsymbol{\rho}^{\mu}, 
\end{aligned}
\end{equation}
where $\mathcal{L}_i$ with $i=N,~K,~l$ represent the lagrangian densities of nucleonic matter, antikaon condensation and leptonic matter respectively.
The lagrangian densities are defined as,
\begin{equation}
\begin{aligned}
\mathcal{L}_N & = \sum_{N} \bar{\psi}_N(i\gamma_{\mu} D^{\mu} - m^{*}_N) \psi_N, \\
\mathcal{L}_K & = D^*_\mu \bar{K} D^\mu K - m^{*^2}_K \bar{K} K, \ \ \text{and} \\
\mathcal{L}_l & = \sum_{l} \bar{\psi}_l (i\gamma_{\mu} \partial^{\mu} - m_l)\psi_l.
\end{aligned}
\end{equation}
%
%\begin{equation} \label{rmftlagrangian}
%    \begin{aligned}
%\mathcal{L} & = \sum_{N} \bar{\psi}_N(i\gamma_{\mu} D^{\mu} - m^{*}_N) \psi_N + \sum_{l} \bar{\psi}_l (i\gamma_{\mu} \partial^{\mu} - m_l)\psi_l \\
% & + D^*_\mu \bar{K} D^\mu K - m^{*^2}_K \bar{K} K + \frac{1}{2}(\partial_{\mu}\sigma\partial^{\mu}\sigma - m_{\sigma}^2 \sigma^2) \\ 
% & -  \frac{1}{4}\omega_{\mu\nu}\omega^{\mu\nu} + \frac{1}{2}m_{\omega}^2\omega_{\mu}\omega^{\mu} - \frac{1}{4}\boldsymbol{\rho}_{\mu\nu} \cdot \boldsymbol{\rho}^{\mu\nu} + \frac{1}{2}m_{\rho}^2\boldsymbol{\rho}_{\mu} \cdot \boldsymbol{\rho}^{\mu} 
%    \end{aligned}
%\end{equation}
Here,  the fields $\psi_N$, $K$ and $\psi_l$ correspond to the baryon, kaon and lepton fields with their bare masses $m_N$, $m_K$ and $m_l$ respectively. The field strength tensors for the scalar meson in Eq.\eqref{rmftlagrangian} is $\sigma$ and vector fields  are $\omega_{\mu \nu}  = \partial_{\mu}\omega_{\nu} - \partial_{\nu}\omega_{\mu}$ and $\boldsymbol{\rho}_{\mu \nu}  = \partial_{\nu}
\boldsymbol{\rho}_{\mu} - \partial_{\mu}\boldsymbol{\rho}_{\nu}$.
The covariant derivative is $D_\mu = \partial_\mu + ig_{\omega j} \omega_\mu + ig_{\rho j} \boldsymbol{\tau}_j \cdot \boldsymbol{\rho}_{\mu}$, with `$j$' denoting the nucleons and antikaons.
We consider the isospin projections as, $\tau_j\equiv \pm1/2$ for $p,~n$ and $\pm1/2$ for $\bar{K}^0,~K^-$ respectively.
The isospin doublets for kaons are denoted by $K\equiv (K^+,K^0)$ and that for antikaons by, $\bar{K} \equiv (K^-, \bar{K}^0)$. The effective nucleon (Dirac) and antikaon masses in the mean-field approximation are given by
\begin{equation} \label{eqn.6}
\begin{aligned}
    m_{N}^* = m_N - g_{\sigma N}\sigma, \quad m_{K}^* = m_K - g_{\sigma K}\sigma
\end{aligned}
\end{equation}
%\sout{where $m_N$, $m_K$ are the bare nucleon and kaon masses respectively.} 

The interactions between mesons, baryons, and antikaons are expressed through the couplings denoted as $g_{ij}$, where $i$ encompasses the mesons and $j$ encompasses the baryons and antikaons. The details of the ground state expectation values for various mesons can be found in \cite{Thapa_2021, Vivek_2020}. The scalar and baryon (vector) number densities are defined as follows for the baryons: $\rho_{N}^s= \langle\bar{\psi}_N \psi_N \rangle$ for scalar density and $\rho_{N}=\langle\bar{\psi}_N \gamma^0 \psi_N\rangle$ for baryon (vector) density, respectively. For the density-dependent model, the coupling constants between mesons and nucleons are density dependent and vary with density as \cite{Malik_2022, Malik_2022_1, Mikhail_2023}:
\begin{equation}
\label{eq:density_dependence}
g_{i N}(\rho)= g_{i N}(\rho_{0}) f_i(x) \quad \quad \text{for }i=\sigma,\omega
\end{equation}
where, $x=\rho/\rho_0$, $\rho_0$ being the nuclear saturation density and
\begin{equation}
\label{fx}
f_i(x)= exp(-x^{a_i-1})
\end{equation}
for $\sigma$ and $\omega$ mesons and
\begin{equation}
\label{fxrho}
f_i(x)= e^{-a_{\rho}(x-1)}
\end{equation}
for the $\rho$-meson.
%\sout{, the density-dependent coupling constant is given by
\begin{equation}
g_{\rho N}(\rho)= g_{\rho N}(\rho_{0}) e^{-a_{\rho}(x-1)}.
\end{equation}
%\ms{{\Large the expression of chemical potential and rearrangement term should be given here.}}
The chemical potential of the $N$-th nucleon in terms of fermi momentum $p_{F_N}$ is written as
\begin{equation} \label{eq:chem_pot}
\begin{aligned}
    & \mu_{N} = \sqrt{p_{F_N}^2 + m_{N}^{*2}} + g_{\omega N}\omega_{0} + g_{\rho N} \boldsymbol{\tau}_{N3} \rho_{03} + \Sigma^{r},
\end{aligned}
\end{equation}
where, the rearrangement term $\Sigma^{r}$ is introduced to maintain the thermodynamic consistency in case of DDRH model  which is given by
\begin{equation}\label{eqn.22}
\begin{aligned}
\Sigma^{r} & = \sum_{N} \left[ \frac{\partial g_{\omega N}}{\partial \rho}\omega_{0}\rho_{N} - \frac{\partial g_{\sigma N}}{\partial \rho} \sigma \rho_{N}^s + \frac{\partial g_{\rho N}}{\partial \rho} \rho_{03} \boldsymbol{\tau}_{N3} \rho_{N} \right],
\end{aligned}
\end{equation}
where $\rho= \sum_{N} \rho_N$ is the total baryon number density, $\rho_{N}^s$ is the scalar density, and $\sigma$, $\omega_0$ and $\rho_{03}$ are the ground state expectation values of various mesonic fields \cite{Thapa_2021}. This re-arrangement term contributes  explicitly only to the matter pressure.

%\ms{{\Large I suggest, the following para should be given at beginning of results or bayesian analysis - }}

The meson-antikaon couplings in the DDRH model are regarded as not being density-dependent \cite{2014PhRvC..90a5801C}. The vector coupling parameters within the kaon sector are determined through the iso-spin counting rule and quark model \cite{2001PhRvC..64e5805B, Pons_2000}, represented as:
\begin{equation}
    g_{\omega K} = \frac{1}{3} g_{\omega N}, \quad  g_{\rho K} = g_{\rho N}
\end{equation}
and for the scalar coupling parameters, they are calculated at nuclear saturation density from the real part of $K^-$ optical potential depth as \cite{Pons_2000}
\begin{equation} \label{ukp_connect}
    U_{\bar{K}} (\rho_0) = - g_{\sigma K} \sigma (\rho_0) - g_{\omega K} \omega_0 (\rho_0) + \Sigma^r_{N} (\rho_0)
\end{equation}
where, $\Sigma^r_{N} (\rho_0)$ is the contribution from nucleons alone. 
%\textcolor{black}{It should be noted that the kaon-meson couplings are directly related to their corresponding vector meson-nucleon couplings, which explicitly exhibit density dependence. As a result, the kaon-meson couplings are also density-dependent, albeit implicitly.}

In case of the $s$-wave antikaons condensates, the number density is given by \cite{1999PhRvC..60b5803G}
\begin{equation} \label{eqn.11}
\begin{aligned}
    \rho_{K^-, \bar{K}^0} & = 2 \left( \omega_{\bar{K}} + g_{\omega K} \omega_0 \pm \frac{1}{2} g_{\rho K} \rho_{03} \right) \\
    & = 2 m^*_K \bar{K} K
\end{aligned}
\end{equation}
%The in-medium energies of $\bar{K}\equiv (K^-,\bar{K}^0)$ for s-wave condensation are provided by
%\begin{equation} \label{eqn.7}
%    \omega_{K^{-} , \bar{K}^0} = m^*_K - g_{\omega K} \omega_0 \mp \frac{1}{2} g_{\rho K} \rho_{03}
%\end{equation}
%with the isospin projections for $K^-,\bar{K}^0$ being $-1/2,+1/2$ respectively. \ms{{\Large Is the equation \eqref{eqn.7} necessary to give? If not delete this.}}

%The expressions for the chemical potential, equation of motion, total energy density ($\varepsilon_f$) and pressure  ($P$) can be found in \cite{Vivek_2020}. The energy density contribution to the total energy from antikaon condensates can be expressed as \begin{equation} \label{eqn.21}
%\varepsilon_{\bar{K}} = m^*_K (\rho_{K^-} + \rho_{\bar{K}^0}),
%\end{equation} 
The total energy density is given by $\varepsilon = \varepsilon_f + \varepsilon_{\bar{K}}$ where, the energy contribution from fermionic matter is given by,
\begin{widetext}
\begin{eqnarray} 
\begin{aligned}
\varepsilon_f & = \frac{1}{2}m_{\sigma}^2 \sigma^{2} + \frac{1}{2} m_{\omega}^2 \omega_{0}^2 + \frac{1}{2}m_{\rho}^2 \rho_{03}^2 + \sum_N \frac{1}{\pi^2} \left[ p_{{F}_N} E^3_{F_N} - \frac{m_{N}^{*2}}{8} \left( p_{{F}_N} E_{F_N} + m_{N}^{*2} \ln \left( \frac{p_{{F}_N} + E_{F_N}}{m_{N}^{*}} \right) \right) \right] \\
	 & + \frac{1}{\pi^2}\sum_l \left[ p_{{F}_l} E^3_{F_l} - \frac{m_{l}^{2}}{8} \left( p_{{F}_l} E_{F_l} + m_{l}^{2} \ln \left( \frac{p_{{F}_l} + E_{F_l}}{m_{l}} \right) \right) \right].
\end{aligned}
\end{eqnarray}
\end{widetext}
Here, $p_{{F}_N}$ and $E_{F_N}$ are the Fermi momentum
and fermi energy of the Nth nucleon, respectively.
The contribution from the antikaon condensation is provided by, $\varepsilon_{\bar{K}} = m^*_K (\rho_{K^-} + \rho_{\bar{K}^0})$, where, 
%\sout{where $m^*_K$ represents the effective mass of antikaons and}
$\rho_{K^-}$ and $\rho_{\bar{K}^0}$ denote the densities of negatively charged and neutral antikaons, respectively. %\ms{{\Large is it $\rho$ or $n$?}} 
Since antikaons act as Bose condensates, they do not directly contribute to the overall matter pressure. The matter pressure is linked to the energy density through the thermodynamic relation (Gibbs-Duhem) represented as
\begin{equation}
p_m = \sum_{N} \mu_N \rho_N + \sum_{l} \mu_l \rho_l - \varepsilon_f,
\end{equation}
as there is no pressure contribution from the antikaons. It is noteworthy to mention that because of the density-dependent nature of couplings, an additional (re-arrangement) term appears to maintain thermodynamic consistency \cite{Hofmann_2001}. This additional term contributes to the matter pressure of the system.
%\ms{[}It is noteworthy to mention that because of the density-dependent nature of couplings, an additional (re-arrangement) term appears to maintain thermodynamic consistency \cite{PhysRevC.64.025804}. This additional term contributes to the matter pressure of the system.
%\ms{] \Large{This part should be rewritten after including rearrangement term above where I mentioned.}}
The charge neutrality and $\beta-$equilibrium conditions with antikaon-condensation are written as
\begin{equation} \label{eqn.18}
\begin{aligned}
 Q^h & = \sum_N q_N \rho^h_N - \rho_e - \rho_\mu = 0, \\
 Q^{\bar{K}} & = \sum_N q_N \rho_N^{\bar{K}} - \rho_{K^-} - \rho_e - \rho_\mu = 0
\end{aligned}
\end{equation}
respectively, where $\rho^h_N$ and $\rho^{\bar{K}}_N$ represents the number densities in hadronic and kaon phases respectively, both having the same form and the chemical equilibrium conditions are, 
\begin{equation} \label{eqn.17}
\begin{aligned}
    \mu_n - \mu_p = \omega_{K^-} = \mu_e, \quad \omega_{\bar{K}^0} = 0,
\end{aligned}
\end{equation}
respectively.

The emergence of an antikaon condensate can arise either through a first-order or second-order phase transition from the hadronic phase to the kaon phase, contingent upon the optical potential depths of $K^-$ at nuclear saturation density. If the transition occurs in a first-order manner, a mixed phase is formed where two states coexist: one with pure hadronic matter lacking a condensate and another with the condensate present. In such a scenario, the Gibbs conditions, along with the conservation of global baryon number and charge neutrality, can be applied to ascertain the state of the mixed phase \cite{1999PhRvC..60b5803G,1998PhRvL..81.4564G,1992PhRvD..46.1274G}. The relevant Gibbs conditions can be found in \cite{Vivek_2020, 1999PhRvC..60b5803G, Bandyopadhyay:2021gnp}.

The approach outlined above provides the EOS within the neutron star core, considering the potential for antikaon condensation.
This core EOS must be seamlessly integrated with the EOS of the crust to solve the Tolman-Oppenheimer-Volkoff (TOV) equations accurately. In this study, the outer crust's EOS is adopted from the seminal work of BPS \cite{BPS}, and the transition from the outer crust to the core is accomplished using a polytropic EOS mode:

\begin{equation}
\label{eq::fulleos}
    P_m(\varepsilon)=
\begin{cases}
    P_{BPS}(\varepsilon), & \text{if } \varepsilon_{min} \leq \varepsilon_{outer} \\
    A+B\varepsilon^\frac{4}{3}, & \text{if } \varepsilon_{outer} < \varepsilon \leq \varepsilon_c \\
    P_{DDRH}(\varepsilon), & \text{if } \varepsilon_c < \varepsilon \\
\end{cases}
\end{equation}

%Here, the $\varepsilon_{min}$ is  $1.0317 \times 10^4$ g-cm$^{-3}$ and the  $\varepsilon_{outer}$ is the energy density at the outer-inner crust transition in \cite{BPS}
\textcolor{black}{In this work, we use the standard BPS EOS for the outer crust, where the energy density ranges from $\epsilon_{\text{min}} = 1.0317 \times 10^4 ~ \text{g}  \text{cm}^{-3}$ to $\epsilon_{\text{outer}} = 4.3 \times 10^{11} ~ \text{g}  \text{cm}^{-3}$. For the region between the core and the outer crust, $\epsilon_{\text{outer}} < \epsilon < \epsilon_c$, where $\epsilon_c$ marks the crust-core transition, we employ a polytropic EOS to simplify computation. This approach is widely used in the literature for computationally demanding calculations. The complete EOS is then described by Eq. \eqref{eq::fulleos}, with the parameters $A$ and $B$ selected to ensure a smooth transition at $\epsilon_{\text{outer}}$ and $\epsilon_c$. The inner crust EOS is matched to the core EOS at $\epsilon_c \sim 2.14 \times 10^{14} \text{g} \text{cm}^{-3}$, in line with recent work in Ref. \cite{Chun_2024}.}

\textcolor{black}{Since the outer crust EOS from BPS is a standard calculation, the transition point from the outer crust to the inner crust is relatively well-established. The primary uncertainty in this region arises from the mass excess of neutron-rich nuclei, which is periodically updated based on experimental measurements  \cite{Chamel2008}. However, the inner crust is highly model-dependent, as the formation of neutron gas and clusters in this region can only be estimated using theoretical models \cite{Parmar_1, Parmar_2}. It has been shown that accurately determining the transition point from the inner crust to the core is crucial, as it can introduce a $\approx$ 3\% error in the measurement of the neutron star radius and crust thickness \cite{Gamba_2020}. Given that we fix the inner crust-core transition point, our calculations inherently carry the associated error.
}

\subsection{Bayesian Analysis}

In Bayesian estimations of the parameters, the probability distribution of a set of model parameters \textbf{X}, called the posterior density function (PDF), given data ($D$) is based on Bayes' theorem and is written as,

\begin{equation}
    \label{eq:bays}
     P(\textbf{X}|D)=\frac{P(D|\textbf{X})P(\textbf{X})}{P(D)}.   
\end{equation}
Here, $P(\textbf{X})$ is the prior probability of the parameter set $\textbf{X}$. The prior probability is updated by the experimental or observational data through a likelihood function, $P(D|\textbf{X})$. The quantity $P(D)$ in the denominator is known as the normalization constant to make sure that the posterior distribution $ P(\textbf{X}|D)$ is a true probability distribution. In the present work, the parameter set $\textbf{X}$ is composed of the parameter of the DDRH Lagrangian given by Eq. \eqref{rmftlagrangian}. Therefore, the parameter set $\textbf{X}$ becomes ($g_{\sigma N}$, $g_{\omega N}$, $g_{\rho N}$, $a_\sigma$, $a_\omega$, $a_\rho$, $U_{\bar{K}}$). 
This study employs the nested sampling Monte Carlo algorithm MLFriends \cite{Buchner_2016, Buchner_2019} with the UltraNest\footnote{\url{https://johannesbuchner.github.io/UltraNest/}} package \cite{Buchner2021} for weighted sampling of the parameter vector $X$.
UltraNest handles a wider range of problems. This includes multiple solutions/modes, non-linear correlation among parameters and posteriors with heavy or light tails. The Slice sampler \cite{2019MNRAS.483.2044H} within UltraNest was utilized, as it is particularly adept and effective for sampling in high-dimensional spaces. Additionally, it guarantees consistent convergence speeds during the sampling procedure. 
The number of steps in the slice sampler is determined by running a sequence of nested sampling runs with an increasing number of steps, stopping when natural logarithm of the model evidence (also known as the marginal likelihood log(Z))  converges. 
%\textcolor{red}{write about log(Z)...} converges.

\subsubsection{Parameters and Priors}
The coupling constants of the DDRH models are treated as free parameters in our Bayesian analysis. The prior distribution for these coupling constants is detailed in Table \ref{tab:prior}. The range for the minimum and maximum values of the coupling constants is informed by Ref.\cite{Malik_2022}, where Bayesian inference was conducted on hyperon signatures within neutron stars using a model akin to the one employed in this study. 
In the context of antikaon potential, experimental investigations \cite{LI1997372, 2000NuPhA.674..553P} demonstrate that kaons undergo a repulsive interaction within nuclear matter, while antikaons experience an attractive potential. Various model computations \cite{KOCH19947} yield a wide spectrum of optical potential values, typically falling within the range of -120 $\le$ $U_{\bar{K}}$ $\le$ -40 MeV. Another study employing a hybrid model \cite{Friedman_1999} suggests that the $K^-$ optical potential ranges from -180 $\pm$ 20 MeV at the nuclear saturation density. 
It shows the uncertainty in the antikaon optical potential at the saturation density. Particularly, it was argued that the impact of $\Lambda (1405)$ resonance state on the antikaon potential led to the small attractive antikaon potential in normal nuclear matter. 
In this study, we have opted for a wide $K^-$ potential range of -180 $\le$ $U_{\bar{K}}$ $\le$ 0 MeV, in order to get a better estimation of the antikaon potential which is the main aim of the present work.

\begin{table}[]
\caption{The prior ($P$) configuration used for the parameters of the DDRH model in this study. The terms 'min' and 'max' refer to the lower and upper limits of the considered distribution, respectively.}
\label{tab:prior}
\begin{tabular}{@{}llll@{}}
\toprule
\toprule

Parameters     & \multicolumn{1}{c}{Prior} & Minimum & Maximum \\ \midrule
$M_{\sigma N}$ (MeV) & Fixed                   & \multicolumn{1}{c}{550}     & \multicolumn{1}{c}{550}    \\
$M_{\omega N}$  (MeV)& Fixed                   & \multicolumn{1}{c}{783}     & \multicolumn{1}{c}{783}    \\
$M_{\rho N}$ (MeV)& Fixed                   & \multicolumn{1}{c}{783}     & \multicolumn{1}{c}{783}    \\

$g_{\sigma N}$ & Uniform                   & \multicolumn{1}{c}{8.5}     & \multicolumn{1}{c}{12}    \\
$g_{\omega N}$ & Uniform                   & \multicolumn{1}{c}{9.5}     & \multicolumn{1}{c}{14}    \\
$g_{\rho N}$   & Uniform                   & \multicolumn{1}{c}{2.5}     & \multicolumn{1}{c}{8.0}     \\
$a_\sigma$     & Uniform                   & \multicolumn{1}{c}{0.0}     & \multicolumn{1}{c}{0.20}    \\
$a_\omega$     & Uniform                   & \multicolumn{1}{c}{0.0}     &\multicolumn{1}{c} {0.20}    \\
$a_\rho$       & Uniform                   & \multicolumn{1}{c}{0.0}     & \multicolumn{1}{c}{1.0}     \\
$U_{\bar{K}}$    & Uniform                   & \multicolumn{1}{c}{-180}    &\multicolumn{1}{c}{0}       \\ \bottomrule
\end{tabular}
\end{table}

\subsubsection{Constrants}
\textbf{Nuclear matter saturation properties:}
The parameters from the DDRH model are directly linked to the properties of nuclear saturation. Given a specific set of isoscalar and isovector parameters, it becomes possible to calculate several crucial nuclear saturation properties.
The EOS of nuclear matter can be decomposed into two
parts as \cite{malik_2020, Parmar_3}
\begin{equation}
    \epsilon(\rho, \alpha)=\epsilon(\rho, 0) + S(\rho)\alpha^2,
\end{equation}
where $\epsilon$ is the energy per nucleon at a given density $\rho$ and
isospin asymmetry $\alpha=\frac{\rho_n -\rho_p}{\rho_n +\rho_p}$. %\ms{{\Large Is not it $1$?, is it correct definition of $\alpha$?}}. $(\rho)$ \ms{{\Large(Is it $S(\rho)$?)}} 
$S(\rho)$ is defined as the density dependent symmetry energy of the system:
\begin{equation}
    S(\rho)=\frac{1}{2}\Big( \frac{\partial ^2 \epsilon(\rho, \alpha)}{\partial \alpha^2}  \Big)_\alpha=0.
\end{equation} 
%\ms{{\Large what is $\delta$? Is it actually $\alpha$ ?}}

$S(\rho)$ at saturation is one of the most crucial nuclear matter properties and is generally denoted as $J$ or $J_{sym,0}$, known as symmetry energy at saturation. As prescribed in \cite{1999PhRvC..60b5803G}, the EOS can be expressed using different bulk nuclear matter properties at the saturation density. Specifically, for symmetric nuclear matter, these properties include the energy per nucleon $\epsilon_0 = \epsilon(\rho_0, 0) (n = 0)$, where $\rho_0$ is the saturation density, the incompressibility coefficient $K_0$ (n = 2), the skewness $Q_0$ (n = 3), and the kurtosis $Z_0$ (n = 4):

\begin{equation}
    X_0^n=3^n\rho_0^n \Big(\frac{\partial^n \epsilon(\rho, 0)}{\partial \rho^n}\Big)_{\rho_0}; \hspace{1cm} n=2,3,4
\end{equation}

Similarly, the symmetry energy can be expanded as a Taylor series around the saturation density $\rho_0$ and the slope $L_{sym,0}$ (n = 1), the curvature $K_{sym,0}$ (n = 2), the
skewness $Q_{sym,0}$ (n = 3), and the kurtosis $Z_{sym,0}$ (n = 4),
respectively, are defined as

\begin{equation}
    X_{sym,0}^n=3^n\rho_0^n \Big(\frac{\partial^n S(\rho)}{\partial \rho^n}\Big)_{\rho_0}; \hspace{1cm} n=1,2,3,4.
\end{equation}

These properties of nuclear matter saturation are reasonably well-constrained by experimental data, providing a known plausible range for some of these values. The table \ref{tab:cosntraints} shows the constraints used in the present work.
%\ms{{\Large $J_{sym,0}$ appeared in table \ref{tab:cosntraints} is not defined}}

For a value of certain nuclear matter property (NMP) at the saturation density, we use a probability function ($p_{NMP}$) as follows \cite{Chun_2024}: 
\begin{equation}
\label{eq:nmp_pfunc}
\begin{aligned}
\text{center} &= \frac{NMP_{\text{low}} + NMP_{\text{up}}}{2}, \\
\text{width} &= \frac{NMP_{\text{up}} - NMP_{\text{low}}}{2}, \\
p_{NMP} &= -0.5 \times \frac{\left| \text{center} - NMP \right|^{10}}{\text{width}^{10}}.
\end{aligned}
\end{equation}
Here, $low$ and $up$ represents the lower and upper range of the given nuclear matter property. Such a function is  a super-Gaussian function and is less extreme than a hard
cut, but strongly disfavours values outside the nominal range.
Furthermore, it significantly helps in the convergence speed of the inference \cite{Chun_2024}.
%\ms{{\Large I think, $p$ and $P$ are same. Make them uniform.}}

%\ms{{\Large What is the meaning of $N^3LO$ in the table \ref{tab:cosntraints}?}}
\begin{table}[]
\caption{The constraints used in the Bayesian inference of the
model parameters to generate the DDRH set with antikaons. These are saturation density $\rho_0$, binding energy $e_0$, incompressibility $K_0$, symmetry energy $J_{sym,0}$ and energy per particle derived from the  $N^3LO$ calculation  based on chiral nucleon-nucleon
(NN) and three-nucleon (3N) interactions for the symmetric (SNM) and pure neutron matter (PNM) \cite{Drischler_2016}. The $N^3LO$ band is expanded by 5\%.}
\label{tab:cosntraints}
\begin{tabular}{@{}cllll@{}}
\toprule
\toprule

\multicolumn{5}{c}{Nuclear Matter  Constraint}                               \\ \midrule
\multicolumn{1}{l}{}    & Parameter    & Unit          & Value/Band    & Ref \\ \midrule
\multirow{5}{*}{SNM}    & $\rho_0$     & $fm^{-3}$     & 0.153 $\pm$ 0.005 &  \cite{TYPEL1999331}   \\
                        & $\epsilon_0$ & MeV         & -16.1 $\pm$ 0.2   &  \cite{Dutra_2014}   \\
                        & $K_0$        & MeV         & 230 $\pm$ 40      &   \cite{Rutel_2005}  \\
                        & $J_{sym,0}$  & MeV         & 30 $-$ 35    & \cite{GARG200736, PhysRevC.76.051603, PhysRevLett.94.032701,Essick_104}  \\
                        & $\varepsilon(\rho)/\rho$    & MeV & $N^3LO$        &  \cite{Drischler_2016}   \\
                        &              &               &                &   \\
\multicolumn{1}{l}{PNM} & $\varepsilon(\rho)/\rho$    & MeV  & $N^3LO$       &  \cite{Drischler_2016}    \\ \bottomrule
\end{tabular}
\end{table}

\textbf{Symmetric (SNM) and pure neutron (PNM) matter:} In addition to the NMP, we also use the constraint from the the chiral EFT calculations for SNM and
PNM by Drischler \textit{et al.} \cite{Drischler_2016}. The authors employed many-body perturbation theory with seven different Hamiltonians incorporating chiral NN and three-nucleon interactions to calculate the energy per particle for various isospin values at low density.  \textcolor{black}{We use the same distribution as in Eq. \eqref{eq:nmp_pfunc}. Since the data from the EFT calculation is extracted manually, energy values at densities ranging from 0.02 to 0.2 fm$^{-3}$ (as provided in Drischler \textit{et al}) \cite{Drischler_2016} are sampled at ten equidistant point}. Additionally, $\chi$EFT predicted bands are expanded by 5\% to appropriately incorporate other relevant \textit{ab initio} calculations. We have not used the constraint on the pressure due to the uncertainty propagation in taking the density derivative of the energy per nucleon as prescribed in \cite{Carreau_2019}.

\textbf{GW170817:} The likelihood for GW170817 is computed by accurately interpolating the likelihood presented in \cite{Hernandez_2020}, which was derived from fitting the strain data released by LIGO/Virgo and is contained within the python package toast\footnote{\url{https://git.ligo.org/francisco.hernandez/toast}}. It is denoted as 

\begin{equation}
    \mathcal{L}_{GW170817}= F(\Lambda_1, \Lambda_2, \mathcal{M},q), 
\end{equation}

where the chirp mass $\mathcal{M}$ is calculated as $\mathcal{M} = (M_1 M_2)^{(3/5)}/(M_1 + M_2)^{(1/5)}$, where $M_1$ and $M_2$ are the masses of the binary components and $q$ represents their mass ratio $(M_1/M_2)$. The tidal deformabilities $\Lambda_1(M_1)$ and $\Lambda_2(M_2)$ signify the mass-dependent deformability of each star. To determine the tidal deformability, mass, and radius of a star, one must solve the perturbed tidal field equation \cite{2008ApJ...677.1216H} alongside the TOV equation simultaneously. This involves integrating both equations from the center of the star to its surface, where the pressure becomes negligible, upon establishing the EOS and the central pressure of the star.

\textbf{PSR J0030+0451 and PSR J0740+6620:} The mass and radius measurements of two pulsars, PSR J0030+0451 and PSR J0740+6620, estimated by the NICER collaborations, have provided stringent constraints on the EOS. For PSR J0030+0451, Riley \textit{et al.} \cite{Riley_2019} reported, at a 68\% confidence level, mass and radius values of M = $1.34^{(+0.15)}_{(-0.16)}$ $M_\odot$ and R = $12.71^{(+1.14)}_{(-1.19)}$ km, or M = $1.44^{(+0.15)}_{(-0.14)}$ $M_\odot$ and R = $13.02^{(+1.24)}_{(-1.06)}$ km  by Miller \textit{et al.} \cite{Miller_2019}. Similarly, for PSR J0740+6620, Miller \textit{et al.} \cite{Miller_2021} found M = $2.072^{(+0.067)}_{(-0.066)}$ $M_\odot$ and R = $12.39^{(+1.30)}_{(-0.98)}$ km, or M = $2.062^{(+0.090)}_{(-0.091)}$ $M_\odot$ and R = $13.71^{(+2.61)}_{(-1.50)}$ km \cite{Miller_2021}.

We incorporated the latest mass measurement of the heavy-mass pulsar PSR J0740+6620, obtained through radio timing (2.08 $\pm$ 0.07 M$_\odot$) \cite{2021ApJ...915L..12F}, along with the
ST+PST (Single-Temperature+Protruding-Single-Temperature) model samples for PSR J0030+0451 (Riley \textit{et al.}, \cite{riley_2019_3386449}) and the NICER and XMM samples for PSR J0740+6620 (Riley \textit{et al.},  \cite{riley_2021_4697625}) in conjunction with KDE method to generate posterior distributions, which are treated as likelihood in our analysis as prescribed in \cite{Zhu_2023, Chun_2024}.

Incorporating data from NMP, gravitational waves (GW), and NICER mass-radius measurements, we adopt the following total likelihood function form:
\begin{equation}
    \mathcal{L}(D|X) = \mathcal{L}_{NMP} \times \mathcal{L}_{GW170817} \times \mathcal{L}_{NICER}.
\end{equation}

\begin{figure*}
    \centering
    \includegraphics[scale=0.3]{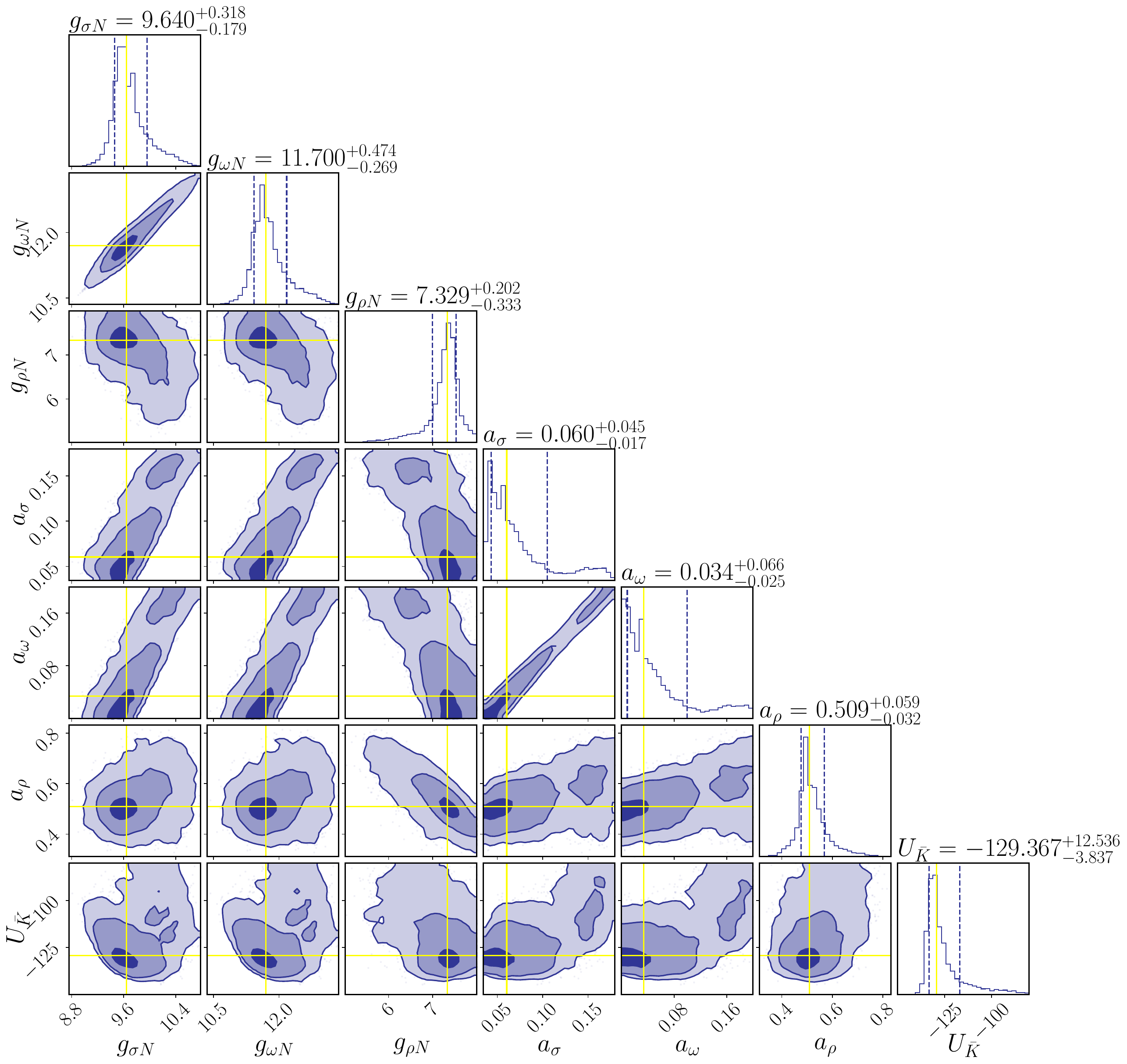}

    \caption{The marginalized posterior distributions of the model parameters. Vertical lines mark the 68\% confidence intervals (CIs). Additionally, the plot includes ellipses representing the 1$\sigma$, 2$\sigma$, and 3$\sigma$ CIs, with darker shades indicating tighter confidence intervals and lighter shades indicating wider intervals in the two-dimensional posterior distributions.}
    \label{fig:par_post}
\end{figure*}

\begin{table*}[]
\centering
\caption{Comparison of the present median values and the associated 68\% and 90\% CIs and previously available values/range of antikaon Potential $U_{\bar{K}}$}
\label{tab:anti-kaon_potentials}
\resizebox{0.9\textwidth}{!}{%
\begin{tabular}{@{}c@{\hskip 0.7cm}c@{\hskip 0.7cm}c@{\hskip 0.7cm}c@{}}
\toprule
\toprule
\multicolumn{1}{l}{} &
  Method &
  \multicolumn{2}{c}{(Anti)kaon Potential (MeV)} \\
  \midrule
\multirow{2}{*}{Present work} &
  \multirow{2}{*}{\begin{tabular}[c]{@{}c@{}}Bayesian analysis using the constraints on  nuclear matter \\ and neutron star observables\end{tabular}} &
  68\% CI &
  90\% CI \\
 &
   &
  $-129.36^{+12.536}_{-3.837}$ &
  $-129.36^{+32.617}_{-5.696}$ \\
\multicolumn{4}{l}{} \\
\multirow{5}{*}{Previous studies} &
  Analysis of the scattering of $K^-$-proton \cite{KOCH19947} &
  \multicolumn{2}{c}{-120 $\leq U_{\bar{K}} \leq -40$} \\
 &
  \begin{tabular}[c]{@{}c@{}}Fitting relativistic mean field (RMF) potential  with \\$K^-$ atomic data \cite{Friedman_1999}\end{tabular} &
  \multicolumn{2}{c}{-180 $\pm$ 20} \\
 &
  \begin{tabular}[c]{@{}c@{}}Self-consistent calculations based on a chiral lagrangian or \\ meson-exchange potentials \cite{PhysRevC.65.054907, SCHAFFNERBIELICH2000153, Mishra2009}\end{tabular} &
  \multicolumn{2}{c}{-80 $\leq U_{\bar{K}} \leq$ -50} \\
   \bottomrule

\end{tabular}
}
\end{table*}
\section{\label{results} Results}

In this segment, we initially examine the posterior probability distributions concerning our model with antikaon condensation. We discuss the interconnections among input variables, parameters related to the isoscalar and isovector aspects of the nuclear EOS, as well as specific properties of NSs in the context of antikaon condensation in their interiors. We analyze the marginalized PDFs for both input variables and derived parameters of Nuclear Matter (NM), PNM, and NSs. Ultimately, we seek correlations among all these variables and look for similarities or tensions with the available trends in the literature.

Our model for the dense matter EOS is formulated assuming the presence of nucleons ($n$ and $p$) and leptons ($e^-$ and $\mu^-$) with provision for antikaon condensation as density inside the star increases. 
Employing a Bayesian parameter estimation approach, we derive distributions of DDRH model parameters while considering constraints on nuclear matter saturation properties as given in Table \ref{tab:cosntraints} (i.e. saturation density $\rho_0$, binding energy $\epsilon_0$, incompressibility $K_0$, symmetry energy $J_{sym,0}$), EOS for SNM and PNM from N$^3$LO calculations in $\chi$EFT \cite{Drischler_2016}, and relevant astrophysical constraints from PSR J0030+0451 \cite{riley_2019_3386449} and PSR J0740+66 \cite{riley_2021_4697625} for neutron star mass-radius relationships, as well as tidal deformability from GW170817 \footnote{The data from GW170817 are available at \url{https://dcc.ligo.org/LIGO-P1800115/public}} \cite{PhysRevLett.119.161101}. In the present work, we utilize Eq. \eqref{fx} for the density dependence of the mesonic field, which enables us to decrease the number of priors, albeit at the cost of relinquishing some control over the density dependence of couplings, as observed in established models like DDME2 \cite{DDME2} and DD2 \cite{DD2}. \textcolor{black}{The couplings in Refs. \cite{DDME2, DD2} are derived based on boundary conditions and are not independent. Therefore, they cannot be directly used as inputs in the Bayesian inference. Instead, one must start with any two of these constants and deduce the remaining two based on the various boundary conditions specified in. This makes the computation extremely intensive and expensive.}

%The couplings in references \cite{DDME2, DD2} are  , governed by diverse boundary conditions that render the Bayesian analysis computationally intensive.

The EOSs generated are subjected to the condition of causality ($c_s/c<1$), thermodynamic stability ($dP/d\rho>0$), the maximum observed NS mass, M$_{\text{max}}$ $\ge$ 2 M$_\odot$ \cite{2021ApJ...915L..12F} and the positiveness of the symmetry energy at all densities \cite{Abbott_2018}. 
For our purpose, we modify the Inference package \texttt{EoS\_inference}
 \footnote{\url{https://github.com/ChunHuangPhy/EoS_inference}}. 
\textcolor{black}{It should be noted that the saturation density $\rho_0$, which appears in Eq. \eqref{eq:density_dependence}, is calculated self-consistently for each model. If $\rho_0$ falls outside the predefined density range, in this case $0.1 \le \rho_0 \le 0.2$ fm$^{-3}$, the sampler will reject that set of input parameters. A similar approach was taken in Ref. \cite{Mikhail_2023}.} 
Around 11,000 EOS configurations are generated after evaluating $\approx$ 1,50,000 likelihood function, and we use the Python package \textit{corner.py} \footnote{\url{https://corner.readthedocs.io/en/latest/}} to visualize one- and two-dimensional projections of the samples \cite{Foreman-Mackey2016}. In the 2D plots, we delineate contours at 1$\sigma$ (39.3\%), 68\%, and 90\% confidence intervals (CIs).

In Fig. \ref{fig:par_post}, we plot the distribution of the parameter of our model i.e. $g_{\sigma N}$, $g_{\omega N}$, $g_{\rho N}$, $a_{\sigma}$, $a_{\omega}$, $a_{\rho}$ and $U_{\bar{K}}$ where $U_{\bar{K}}$ also acts as an input parameter.  In the corner plot, the diagonal blocks display the marginalized 1D distribution for each individual parameter. Within these distributions, vertical lines represent the 68\% lower bound, median, and 68\% upper bound CIs of the model parameters, the values of which are shown at the top of each parameter. Due to the stringent constraint imposed on binding energy  $\epsilon_0$ and saturation density $\rho_0$  (see Table \ref{tab:cosntraints}), the coupling constants $g_{\sigma N}$ and $g_{\omega N}$ exhibit a high degree of correlation, evident in the highly elliptical shape of their 2D CIs. These coupling constants at the saturation density also display moderate correlation with $a_{\sigma}$ and $a_{\omega}$, which signify the nonlinearity strength in the isoscalar sector. Similarly, $a_{\sigma}$ and $a_{\omega}$ demonstrate a strong correlation. 
The introduction of antikaons notably softens the EOS \cite{Vivek_2020, Glendenning_1999, 2000NuPhA.674..553P, Sarmistha_2001, 2001PhRvC..64e5805B, Sarmistha_2002}. Consequently, lower values of $a_{\sigma}$ and $a_{\omega}$ are preferred, indicating a preference for a stiffer EOS to meet the $2 M_\odot$ constraint. This observation aligns with the findings of Ref. \cite{Malik_2022}, which explores Bayesian inference regarding hyperon signatures within NSs, where hyperons similarly soften the EOS.  
The isovector coupling constant $g_{\rho N}$ exhibits no significant correlation with other parameters. However, it's worth noting that a smaller value of $g_{\rho N}$ is favoured in the presence of antikaon condensate compared to scenarios involving only hadronic matter \cite{Malik_2022_1} or hyperonic matter \cite{Malik_2022}, where it is estimated to be $\sim 4$.
A factor of $\frac{1}{2}$ should be noted in our Lagrangian as compared to the Ref. \cite{Malik_2022, Malik_2022_1}.
%The higher value of $g_{\rho N}$ is consistent with the well known DDRH models DDME2 \cite{DDME2} and DD2 \cite{DD2}. 
Furthermore, We do not get a peak for $g_{\rho N}$ and $a_{\rho}$ without the constraints from $\chi$EFT calculation on SNM and PNM binding energy. The PDFs of the input parameters display a Gaussian-like distribution, indicating that the simulation has converged well. The qualitative behavior of the parameters aligns with the findings of Refs. \cite{Mikhail_2023, Malik_2022_1, Malik_2022}, which also utilize the same CDF formalism for analyzing hadronic and hyperonic EOSs. Lastly and notably, the antikaon potential exhibits no discernible correlation with any model parameter, as evidenced by the nearly circular shape of its 2D CIs.
In Table \ref{tab:anti-kaon_potentials}, we showcase the outcomes of our investigation regarding the antikaon potential, encompassing the 68\% and 90\% CI, along with additional ranges reported in existing literature. Our study provides the most stringent constraints on the antikaon potential, aligning with diverse nuclear matter and astrophysical constraints available.

\begin{figure*}
    \centering
    \includegraphics[scale=0.3]{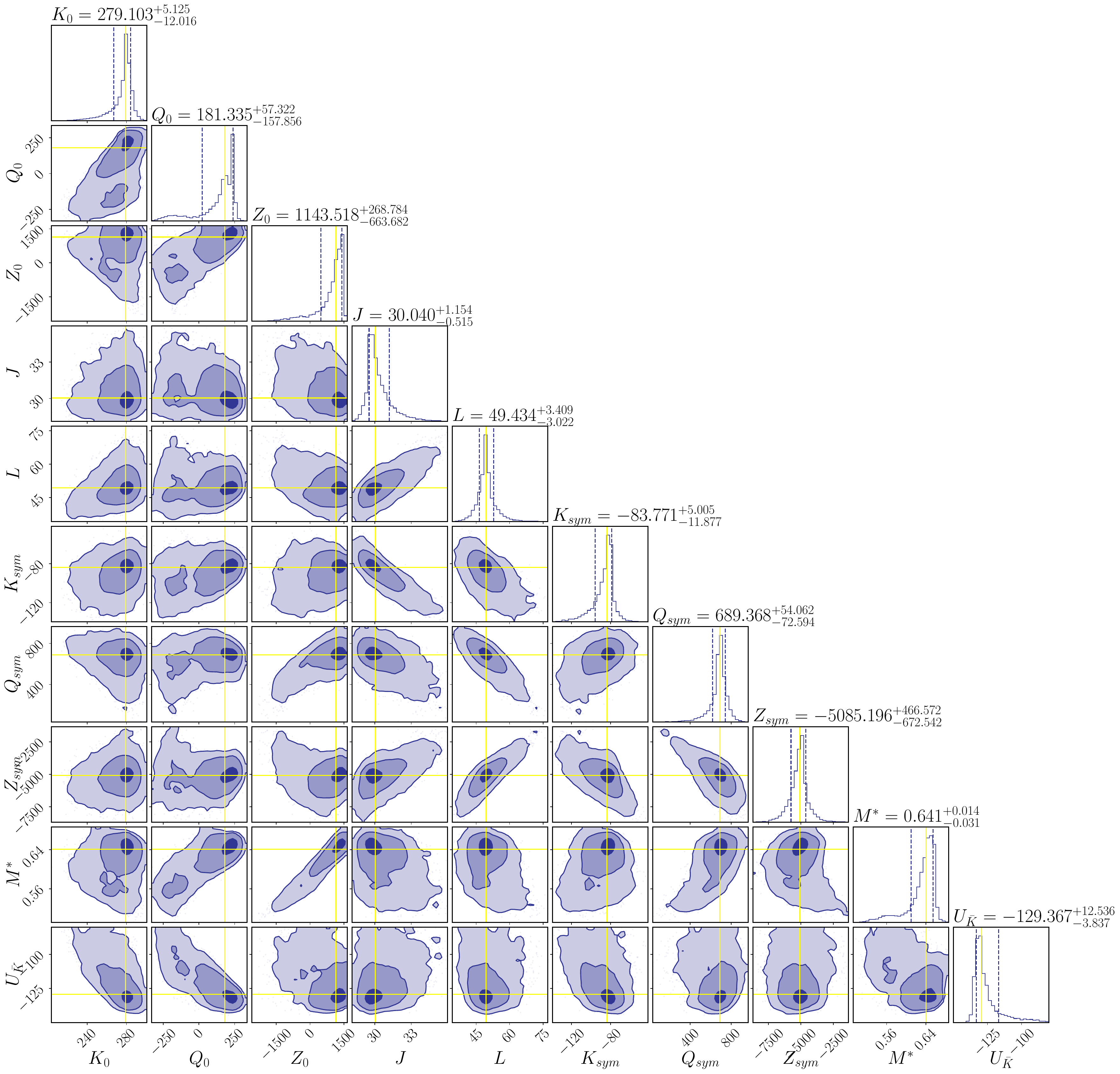}

    \caption{The marginalized posterior distributions of the nuclear matter properties of the parameter sets in Fig. \ref{fig:par_post}. Vertical lines mark the 68\% CIs. Additionally, the plot includes ellipses representing the 1$\sigma$, 2$\sigma$, and 3$\sigma$ CIs, with darker shades indicating tighter confidence intervals and lighter shades indicating wider intervals in the two-dimensional posterior distributions.}
    \label{fig:nuc_prop}
\end{figure*}

\begin{figure*}
    \centering
    \includegraphics[scale=0.3]{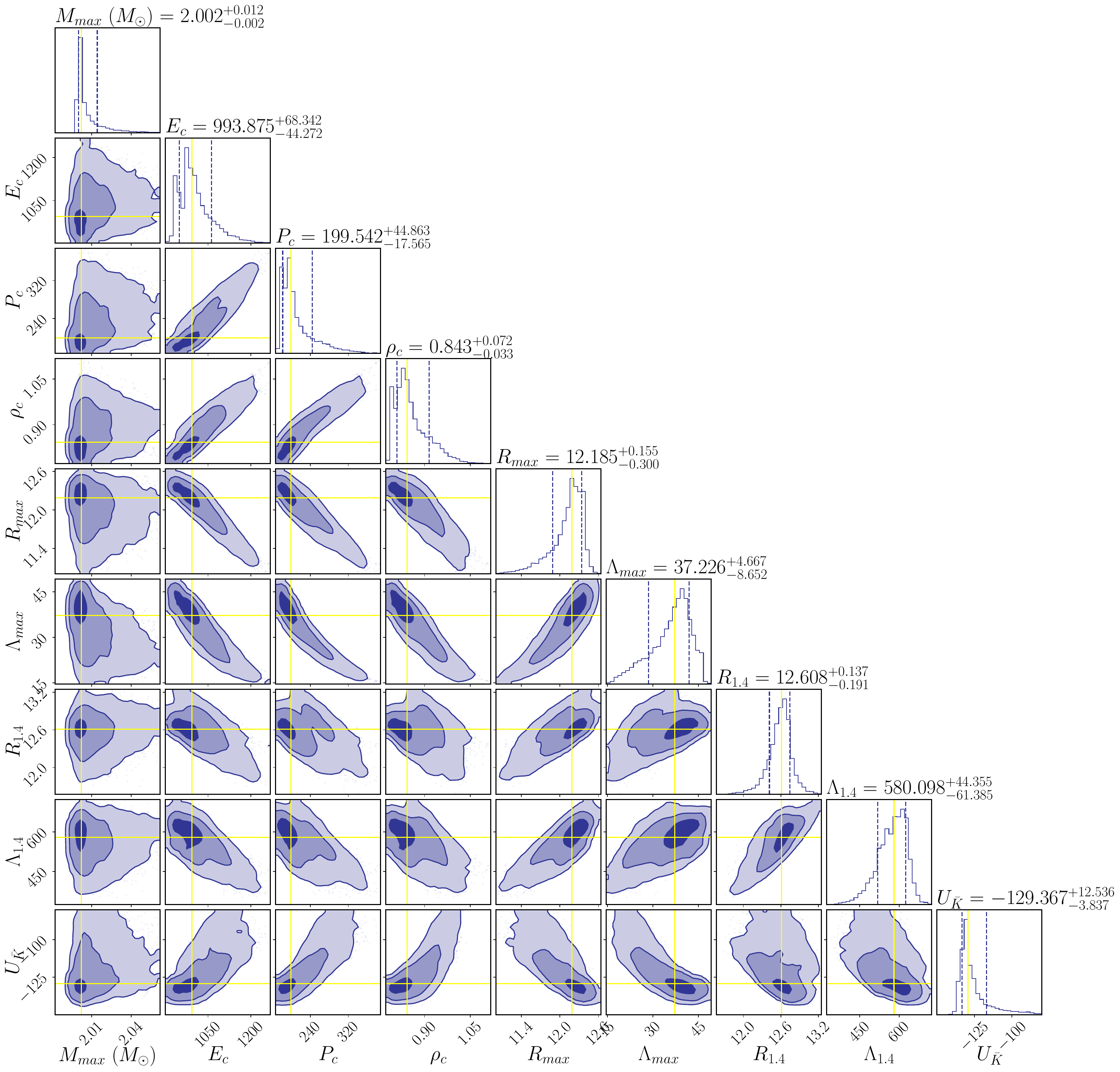}

    \caption{The marginalized posterior distributions of the NS properties namely; Maximum mass ($M_{max}$), central energy density  ($E_c$) in MeV-fm$^{-3}$, central pressure ($P_c$) in MeV-fm$^{-3}$, central baryon density in  fm$^{-3}$ ($\rho_c$), radius corresponding to the  maximum mass ($R_{max}$) in Km, tidal deformability corresponding to the maximum mass $(\Lambda_{max})$, radius of the canonical star $(R_{1.4}$) in Km, and tidal deformability of the canonical star ($\Lambda_{1.4}$) along with the distribution of $U_{\bar{K}}$. The 2D ellipse and vertical lines  have the same meaning as in Fig. \ref{fig:par_post}.}
    \label{fig:tov_post}
\end{figure*}

Next, we conduct a statistical analysis of the nuclear matter properties by utilizing the posterior distributions derived from the calculated model parameters. In Fig. \ref{fig:nuc_prop}, we show the marginalized 2D and 1D posterior distribution of the NMPs. 
The verticle lines represent the 68\% CIs while the 2D ellipses represent the 1$\sigma$, 2$\sigma$, and 3$\sigma$ CIs as in Fig. \ref{fig:par_post}.  While the value of $J$ and $L$ are reasonably within the available range from various experimental analyses such as analysis of giant and pygmy dipole resonances, isospin diffusion measurements, isobaric analog states etc. \cite{GARG200736, PhysRevC.76.051603, PhysRevLett.94.032701}, the compressibility is on the higher side of the currently allowed range measured from the analysis of isoscalar giant monopole resonances in heavy nuclei as $K_0$=240 $\pm$ 10 MeV \cite{PhysRevC.70.024307}, or $K_0$=248 $\pm$ 8 MeV \cite{PhysRevC.69.041301}. 
The reason for this is that while $K_0$ = 230 $\pm$ 40 MeV is a constraint in Bayesian fitting, lower $K_0$ values cannot support a 2 M$_\odot$ configuration in the presence of an antikaon condensate because the EOS becomes significantly smoother. Similarly, $Q_0$ also falls on the higher side when compared with the posterior analysis of the EOS without the antikaon condensation in Refs. \cite{Malik_2022_1, Mikhail_2023}. Although, one should note that the choice of constraints such as $\chi$EFT calculations and astrophysical constraints are slight different in both the study. 
Several correlations can be observed, including some that are well-documented in the literature, such as $K_{sym}-J$, $K_{sym}-L$, $J-L$, and $Z_{sym}-L$.  A more comprehensive analysis of these correlations using the Kendall rank correlation matrix \cite{KENDALL_1938} will be discussed in a subsequent section. Our values for the three coefficients characterizing the high-density behavior of neutron-rich matter, namely the curvature of the symmetry energy $K_{\text{sym},0}$, the skewness of the symmetry energy $Q_{\text{sym},0}$, and the skewness $Q_0$, fall well within their known bounds. These bounds are typically -400 $\leq$ $K_{\text{sym},0}$ $\leq$ 100 MeV, -200 $\leq$ $Q_{\text{sym},0}$ $\leq$ 800 MeV, and -800 $\leq$ $Q_0$ $\leq$ 400 MeV, primarily established through analyses of terrestrial nuclear experiments and energy density functionals \cite{Cai2017, Tews_2017, Zhang2017, Zhang_2019}. The median effective mass $M^*$ is relatively higher compared to some well-known CDFs like DD2 \cite{DD2} and DDME2 \cite{DDME2}, yet it aligns with the Bayesian analysis in \cite{Mikhail_2023}. This effective mass demonstrates a significant dependency on $Q_0$ and $Z_0$; a characteristic also observed in \cite{Mikhail_2023}, indicating a potentially universal feature.

The antikaon potential exhibits no significant correlation with nuclear matter properties at the saturation density, except for the compressibility $K_0$ and skewness $Q_0$. Both of these relationships exhibit a negative linear correlation, indicating that a stiffer EOS at the saturation density (at higher density, EOS become soft due to the onset of antikaon condensate) is preferred for a deeper antikaon potential.
This could be attributed to the fact that, under the current assumptions, the antikaon potential is linked to the isoscalar couplings via Eq. \eqref{ukp_connect}, which in turn heavily influences the compressibility of the EOS. Moreover, since the onset of antikaon condensation occurs at a much higher density, it may not yield a strong correlation with nuclear matter properties at the saturation density.

\begin{figure}
    \centering
    \includegraphics[scale=0.4]{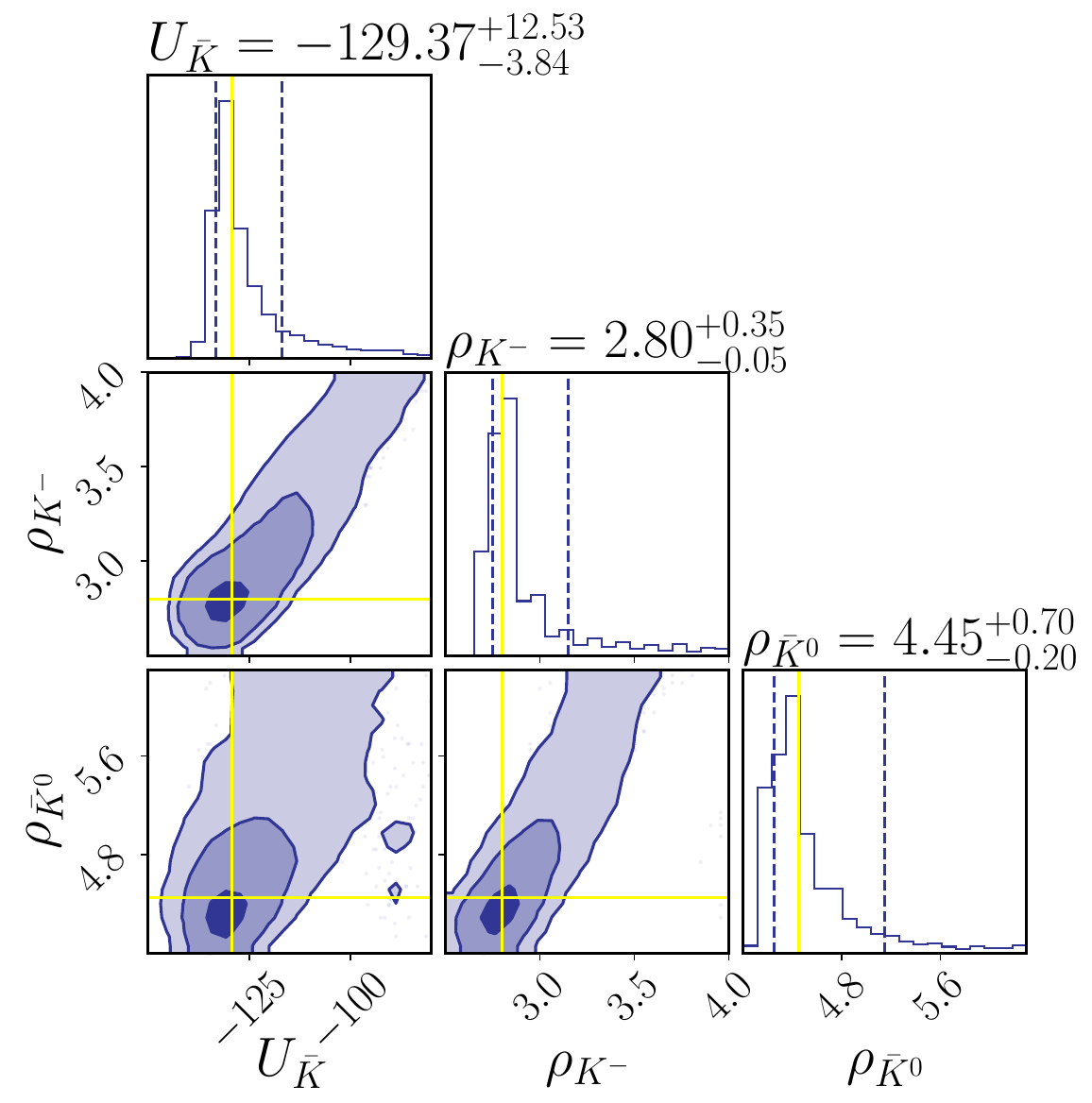}
    \caption{The marginalized posterior distributions of the onset density of the ${K^-}$ and $\bar{K^0}$ as a function of saturation density $\rho_0$. The 2D ellipse and vertical lines  have the same meaning as in Fig. \ref{fig:par_post}. }
    \label{fig:tran_density}
\end{figure}

In Fig. \ref{fig:tov_post}, we present a corner plot showcasing the marginalized posterior distributions 68\%  CI for various neutron star properties, including Maximum mass ($M_{\text{max}}$), central energy density ($E_c$), central pressure ($P_c$), central baryon density ($\rho_c$), radius corresponding to the maximum mass ($R_{\text{max}}$), tidal deformability corresponding to the maximum mass ($\Lambda_{\text{max}}$), radius of the canonical star ($R_{1.4}$), and tidal deformability of the canonical star ($\Lambda_{1.4}$) along with the distribution of $U_{\bar{K}}$. 
The $M_{\text{max}}$ approaches $2 M_\odot$ for the EOS involving antikaon condensates. 
However, the presence of antikaon condensation notably softens the EOS, particularly post the onset of $\bar{K^0}$, which diminishes the likelihood of a significantly large maximum mass. It's worth noting that the maximum mass reported here is lower than that from Bayesian analysis involving hyperons inside neutron stars, as reported in \cite{Malik_2022}, and it exhibits a narrower range. Additionally, the radius of the canonical star ($R_{1.4}$) and the tidal deformability of the canonical star ($\Lambda_{1.4}$) fall well within their respective allowed ranges as derived from NICER \cite{Miller_2021} and GW170817 event \cite{PhysRevLett.119.161101}. 
Recently, S. Huth \textit{et al.} \cite{Huth2022} investigated the constraints on neutron-star matter through microscopic and macroscopic collisions, utilizing data from $\chi$EFT, Multi-messenger astrophysics, and heavy-ion collision (HIC) experiments. 
Our calculated value of $R_{1.4}$ falls comfortably within the reported range of $R_{1.4}= 12.01^{+0.78}_{-0.77}$. The antikaon potential $U_{\bar{K}}$ exhibits a linear correlation with the central energy density ($E_c$) and central pressure ($P_c$) while demonstrating a non-linear relationship with the central baryon density ($\rho_c$). Furthermore, its impact extends to $R_{1.4}$ and $\Lambda_{1.4}$, with the former showing a linear relationship and the latter displaying a non-linear nature.

Fig. \ref{fig:tran_density} shows the posterior distribution of the onset density of the $K^-$ and $\bar{K^0}$ as a function of saturation density $\rho_0$. Both $\rho_{{K^-}}$ and $\rho_{\bar{K^0}}$ are strongly correlated with the depth of antikaon potential, i.e. a deeper antikaon potential favour the early onset of antikaon condensation.  Table \ref{tab:tran_density} shows the median values and the associated 68\% and 90\% CIs of the onset densities, $\rho$ (in units of $\rho_0$) for
antikaon condensation in dense nuclear matter. While the lower bounds of both $\rho_{{K^-}}$ and $\rho_{\bar{K^0}}$ are tightly constrained within their respective CIs of 68\% and 90\%, the upper bounds exhibit significant deviations. This discrepancy is notable when comparing the upper limits with their corresponding CIs. 
Furthermore, the onset density $\rho_{K^-}$ is observed to be more tightly constrained compared to $\rho_{\bar{K^0}}$ within the same CIs. When comparing the central baryon density of the posterior samples, denoted as $\rho_c=0.843^{+0.072}_{-0.003}$ fm$^{-3}$ or approximately $5.6\rho_0$ at a 68\% CI, it becomes apparent that both $K^-$ and $\bar{K}^0$ condensate could feasibly exist within a NS of $M_{\text{max}}\ge2M_\odot$. 
However, for a canonical NS ($M=1.4M_\odot$) with a central density of $\rho_{1.4}=0.365^{+0.026}_{-0.008}$ fm$^{-3}$ or approximately $2.4\rho_0$ at a 68\% CI, the possibility of antikaon condensation is deemed unlikely. This implies that the current observations of the canonical NS mass are not sufficient. 
We will require more astrophysical data on heavier NSs from sources like NICER, along with observations of Mass-Radius ($M-R$) relationships using future large-area X-ray telescopes such as eXTP \cite{RevModPhys.88.021001} and STROBE-X \cite{2019arXiv190303035R}. \textcolor{black}{Additionally, with the anticipated large number of binary neutron star (BNS) observations through gravitational waves (GWs), future third-generation GW detectors, such as Cosmic Explorer (CE) and the Einstein Telescope (ET), will enable us to constrain the tidal deformability and, consequently, the equation of state (EoS) of neutron stars (NS) with exceptional precision.} These observations will be crucial in detecting any potential signatures of antikaon condensate. 
\begin{table}
\caption{The median values and the associated 68\% and 90\% CIs of the onset densities, $\rho$ (in units of $\rho_0$ ) for
antikaon condensation in dense nuclear matter}
\label{tab:tran_density}
\scalebox{1.2}{
\begin{tabular}{@{}clclclll@{}}
\toprule
\toprule
\multicolumn{4}{c}{$\rho^c_{{K^-}}$}                    & \multicolumn{4}{c}{$\rho^c_{\bar{K^0}}$}                    \\ \midrule
\multicolumn{2}{c}{68\% CI} & \multicolumn{2}{c}{90\% CI} & \multicolumn{2}{c}{68\% CI} & \multicolumn{2}{l}{90\% CI} \\
\multicolumn{2}{c}{$2.80^{+0.35}_{-0.05}$}        & \multicolumn{2}{c}{$2.80^{+1.00}_{-0.10}$}        & \multicolumn{2}{l}{$4.45^{+0.70}_{-0.20}$}        & \multicolumn{2}{l}{$4.45^{+2.65}_{-0.25}$}        \\ \bottomrule
\end{tabular}
}
\end{table}

\begin{figure}
    \centering
    \includegraphics[scale=0.37]{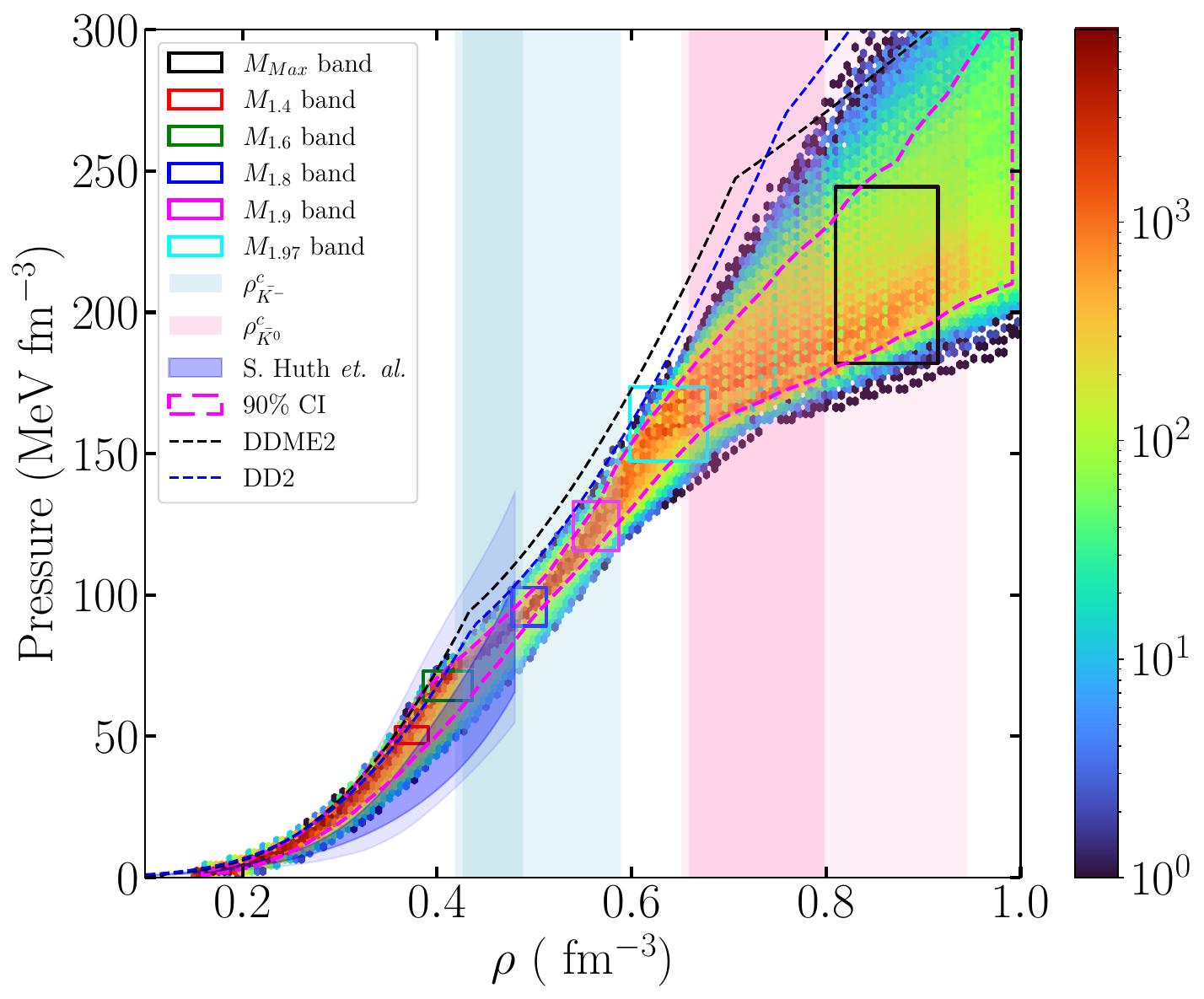}
    \caption{The joint probability distribution $P(P_m,\rho)$ of the Pressure ($P_m$) vs baryon density ($\rho$) and the 90\% CI (magenta dashed line) for the EOS for the neutron star matter with antikaon condensation. The color represents the probability density at a given ($P_m, \rho$). The area represented by the black solid line represents the ($P_m, \rho$) corresponding to the maximum mass as given in Fig. \ref{fig:tov_post}, and the red solid line represents the ($P_m, \rho$) corresponding to canonical mass at 68\% CI.  The EOS is compared with the recent estimation of EOS with the combination of microscopic as well as Multi-messenger astrophysics \cite{Huth2022}. The onset density of $K^{-}$ and $K^{0}$ is shown by the vertical shaded area with 90\% and 68\% CI.
The black and blue dashed line corresponds to the EOS using the model DDME2 and DD2 at the median $U_{\bar{K}}$ in Fig. \ref{fig:par_post}.}
    \label{fig:eos}
\end{figure}

We now discuss the posterior of the EOSs obtained in our Bayesian analysis. In Fig. \ref{fig:eos}, we show the joint probability distribution $P(P_m,\rho)$ of the pressure ($P_m$) vs baryon density ($\rho$) and the 90\% CI (magenta dashed line) for the EOS for the neutron star matter with antikaon condensation. The colorbar represents the probability density at a given  ($P_m, \rho$) on log scale. The black solid line represents the central ($P_m, \rho$) corresponding to the maximum mass as given in Fig. \ref{fig:tov_post}, and the red, green, blue, magenta and cyan solid line represents the central ($P_m, \rho$) corresponding to canonical mass, 1.6, 1.8, 1.9 and 1.97 $M_\odot$ at 68\% CI.  The EOS is compared with the recent estimation of EOS with the combination of microscopic as well as Multi-messenger astrophysics \cite{Huth2022}. 
Two distinct kinks can be observed in the EOSs, symbolizing the onset of ${K^-}$ and $\bar{K^0}$, respectively. The 68\% and 90\% CI ranges of the antikaon onset densities $\rho^c_{{K^-}}$ and $\rho^c_{\bar{K^0}}$ are indicated by vertical bands. It is evident that antikaon condensation is not supported within canonical neutron stars, as the onset densities lie outside the central density of such stars. While the condensation of $K^-$ appears feasible for higher neutron star masses ($M>1.6M_\odot$), $\bar{K}^0$ does not appear even for $M=1.97M_\odot$ and appears for $M \ge 2.0M_\odot$. Furthermore, it is observed that the variability in the range of $\rho^c_{{K^-}}$ is lower when compared to $\rho^c_{\bar{K}^0}$. 

It has been observed that lower densities are notably influenced by HIC results, in conjunction with the nuclear matter parameter at the saturation density, contributing significantly to the overall posterior distribution of the EOS. Conversely, the EOS at higher densities is primarily determined by astrophysical observations. 
Since antikaon condensation is not favoured in NSs with masses $M \le 1.4 M_\odot$ as evident in Fig. \ref{fig:eos}, the corresponding EOS should adhere to astrophysical constraints. However, it is important to note that the antikaon potential is governed by Eq. \eqref{ukp_connect}, where the scalar coupling parameter in the kaonic sector is calculated at $\rho_0$. Therefore, to determine the optimal antikaon potential, one requires information from both the nuclear matter parameter at the saturation density and astrophysical observations.
Moreover, upon comparison with the posterior results from S. Huth \textit{et al.} \cite{Huth2022}, where final EOS constraints are obtained through the combination of both the HIC information and astrophysical multi-messenger observations,  our EOS demonstrates reasonable agreement with their findings. 
Although we did not utilize HIC data as a constraint in our study, our EOS aligns well with their results. Furthermore, considering the  the antikaon condensation inside the neutron star, the EOS is relatively stiff in the lower densities as compared to the 68\% CI of S. Huth \textit{et al.} \cite{Huth2022}, which is also reflected from the median value of incompressibility in  Fig. \ref{fig:nuc_prop}. 
We also compare our posterior of the EOS with the two well-known CDF models, DDME2 \cite{DDME2} and DD2 \cite{DD2}, at the median value of $U_{\bar{K}}$ for reference. Both of these models lie outside our posterior, although their qualitative features such as $K^-$ and $\bar{K}^0$ onset density (see the two kink in the EOS) match those of our posterior.
\begin{figure}
    \centering
    \includegraphics[scale=0.37]{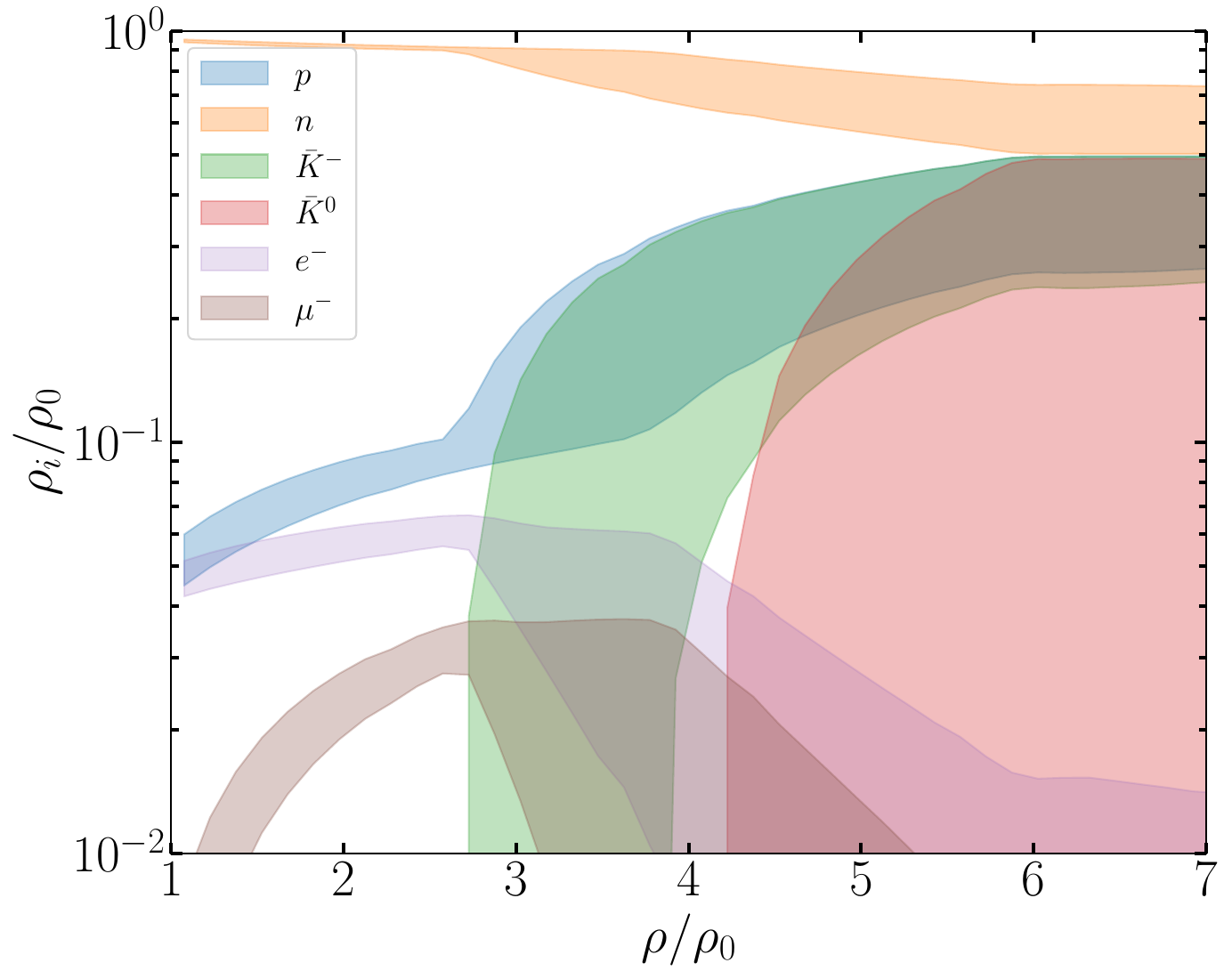}

    \caption{Population densities $\rho_i/\rho$ at 90\% CI obtained from the ensemble of EOS for proton ($p$), neutron ($n$), antikaons $K^-$ and $\bar{K}^0$, electron ($e^-$) and muon ($\mu^-$)  as a function of density (in the unit of $\rho_0$).}
    \label{fig:part_fraction}
\end{figure}

In Fig. \ref{fig:part_fraction}, we examine the population densities $\rho_i/\rho$ for protons ($p$), neutrons ($n$), antikaons $K^-$ and $\bar{K}^0$, electrons ($e^-$), and muons ($\mu^-$) within a 90\% CI. As the negatively charged antikaons emerge, the population density of electrons ($e^-$) and muons ($\mu^-$) starts to decrease. This causes antikaons to condense in the lowest energy state, making them preferable for maintaining the global charge neutrality condition \cite{Vivek_2020}.
The charge neutrality also results in the $K^-$ population density approaching that of protons 
%with the appearance of $\bar{K}^0$ 
and a decrease in the $e^-$ population density. At lower densities ($\leq 2-2.5 \rho_0$), the population densities of various species are more accurately determined compared to higher densities, with $\bar{K}^0$ showing the largest deviation.

\begin{figure}
    \centering
    \includegraphics[scale=0.35]{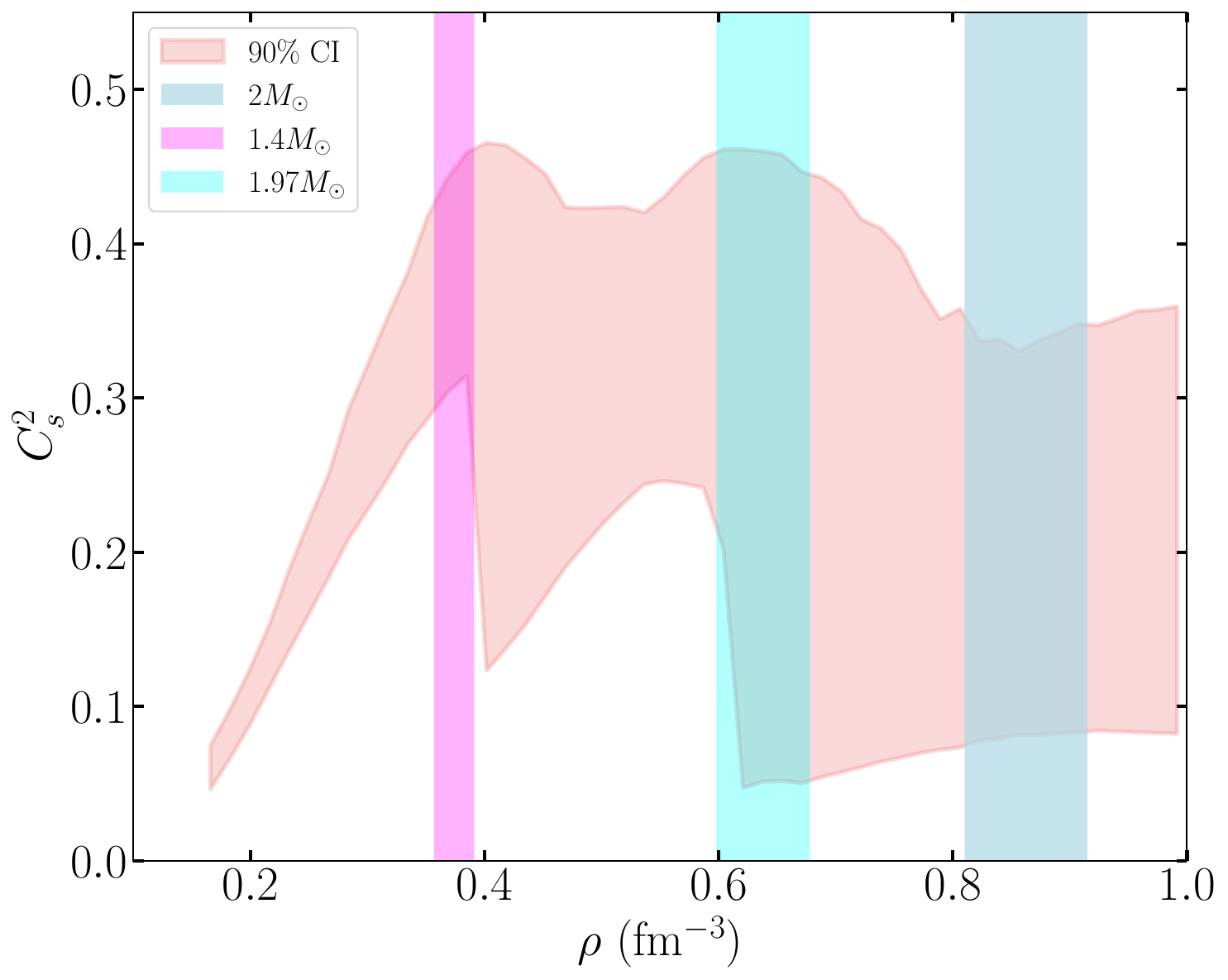}
    \caption{The speed of sound squared ($C_s^2$) as a function of baryon density ($\rho$). The red shaded region indicates the 90\% CI. The vertical shaded regions represent the  central densities of various neutron star masses.}
    \label{fig:sound_speed}
\end{figure}

It was reported in Refs. \cite{PhysRevLett.114.031103, PhysRevD.88.083013, Tews_2018}, and most recently in \cite{Annala2020, Annala2023}, that for all stable NSs to be composed of hadronic matter alone, the EOS must significantly violate the conformal limit $C_s^2 \leq 1/3$ \cite{Cherman_2009}. To analyze the behaviour of the speed of sound in matter composed of antikaon condensation, the speed of sound ($C_s^2 = \frac{dP}{dE}$) is depicted in Fig. \ref{fig:sound_speed} as a function of baryon density. The two kinks represent the appearance of $K^-$ and $\bar{K}^0$ as shown in Fig. \ref{fig:eos}. $C_s^2$ increases with baryon density until the onset of $K^-$ and drops due to the softening of the EOS; this drop is more pronounced for the appearance of $\bar{K}^0$. 
While for the 1.4 and 1.97 $M_\odot$ NSs, the $C_s^2$ is relatively high, it drops significantly for 2 $M_\odot$. This signature of the speed of sound is somewhat similar to the existence of quark-matter cores, where lower values of $C_s^2$ favor the hadron-quark phase transition \cite{Annala2020, Annala2023}. When compared to matter composed solely of hadrons \cite{Malik_2022_1} or matter composed of hyperons \cite{Malik_2022}, the $C_s^2$ shows a significant decrease in its magnitude at high densities. Hence, a significantly low value of $C_s^2$ could serve as a signature of antikaon condensation within NS interiors. Furthermore, the DDRH description of matter used here automatically ensures that the $C_s^2$ is below 1.

\begin{figure}
    \centering
    \includegraphics[scale=0.36]{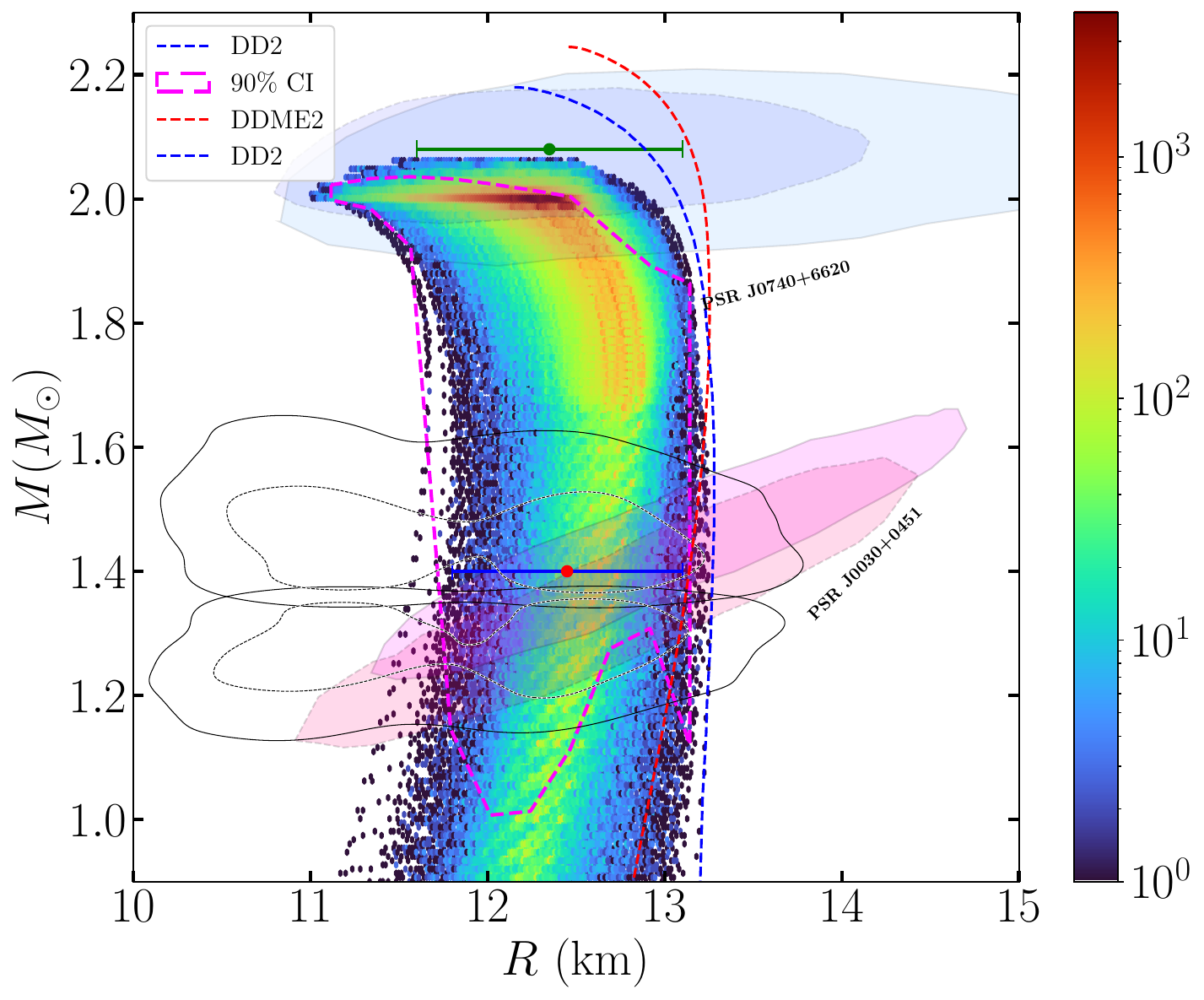}
    \caption{The joint probability distribution $P(M, R)$ for the EOS considering antikaon condensation is shown. The M-R posterior is compared with the 90\% CI (solid lines) and 50\% CI (dashed lines) for the binary components of the GW170817 event, represented by the black lines. The double-headed blue and green lines represent the radius constraints by Miller \textit{et al.} \cite{Miller_2021} for neutron stars of masses 1.4 and 2.08 M$_\odot$, respectively. The shaded regions in dark and light blue represent the constraints from the heavy mass pulsar PSR J0740+6620 by Riley \textit{et al.} \cite{Riley_2019} and Miller \textit{et al.} \cite{Miller_2019}, respectively. Similarly, the shaded regions in pink and orange represent the constraints from the millisecond pulsar PSR J0030+0451 by Riley et al. \cite{riley_2021_4697625} and Miller \textit{et al.} \cite{Miller_2021} The dashed line with magnets represents the 90\% CI region obtained in the present calculation. The black and blue dashed line corresponds to the EOS using the model DDME2 and DD2 at the median $U_{\bar{K}}$ as in Fig. \ref{fig:eos}.} 
    \label{fig:mr}
\end{figure}

We now discuss the NS structure solutions (or, the solutions of TOV equations) for the ensemble of EOSs obtained. In Fig. \ref{fig:mr}, we present the probability distribution $P(M, R)$ for nuclear matter, considering the effect of antikaon condensates. This distribution is compared with various astrophysical constraints, including those from pulsars PSR J0030+0451 and PSR J0740+6620, the binary components of the GW170817 event, and the radius constraints set forth by Miller et al. \cite{Miller_2021} for neutron stars of masses $1.4$ and $2.08$ $M_\odot$, respectively. The color scheme in the plot denotes the probability, transitioning from blue to red as the probability increases. Notably, our M-R posterior aligns remarkably well with the GW170817 (50\% CI) and NICER (PSR J0030+0451) overlap region. The peak probability lies within the range estimated by the GW170817 and NICER data. Furthermore, the M-R posterior, along with its 90\% CI, closely matches the 1.4 M$_\odot$ radius determined by Miller et al. \cite{Miller_2021}, with the central value coinciding with the peak probability of our M-R posterior. It's worth noting that, as demonstrated in our earlier analysis, antikaon condensation is not feasible for stars with masses less than $1.4 M_\odot$. However, for heavier NSs, our results are consistent with the posteriors derived from the massive pulsar, PSR $J0740+6620$. As depicted in Fig. \ref{fig:eos}, we observed that the appearance of $\bar{K}^0$ becomes plausible for NSs with masses $M \geq 2 M_\odot$, resulting in a significantly softer EOS. Consequently, the maximum mass reported in our calculation is approximately $2 M_\odot$, which is lower compared to both pure hadronic EOS \cite{Malik_2022_1, Mikhail_2023} and EOS considering hyperons \cite{Malik_2022}.  When comparing with the M-R profile of the DDME2 and DD2 sets calculated at the median value of $U_{\bar{K}}$, we observe that both of them lie at the extremes of our M-R posterior. Additionally, they estimate larger NS masses. However, DDME2 falls outside the range for $M=2.08M_\odot$ as estimated by \cite{2021ApJ...915L..12F} while DD2 supports it. Evidently, both models lie outside the 90\% confidence interval of  M-R posterior estimated in \cite{Malik_2022_1} which consider only $n$ and $p$ in the system.

\begin{figure*}
    \centering
    \includegraphics[scale=0.35]{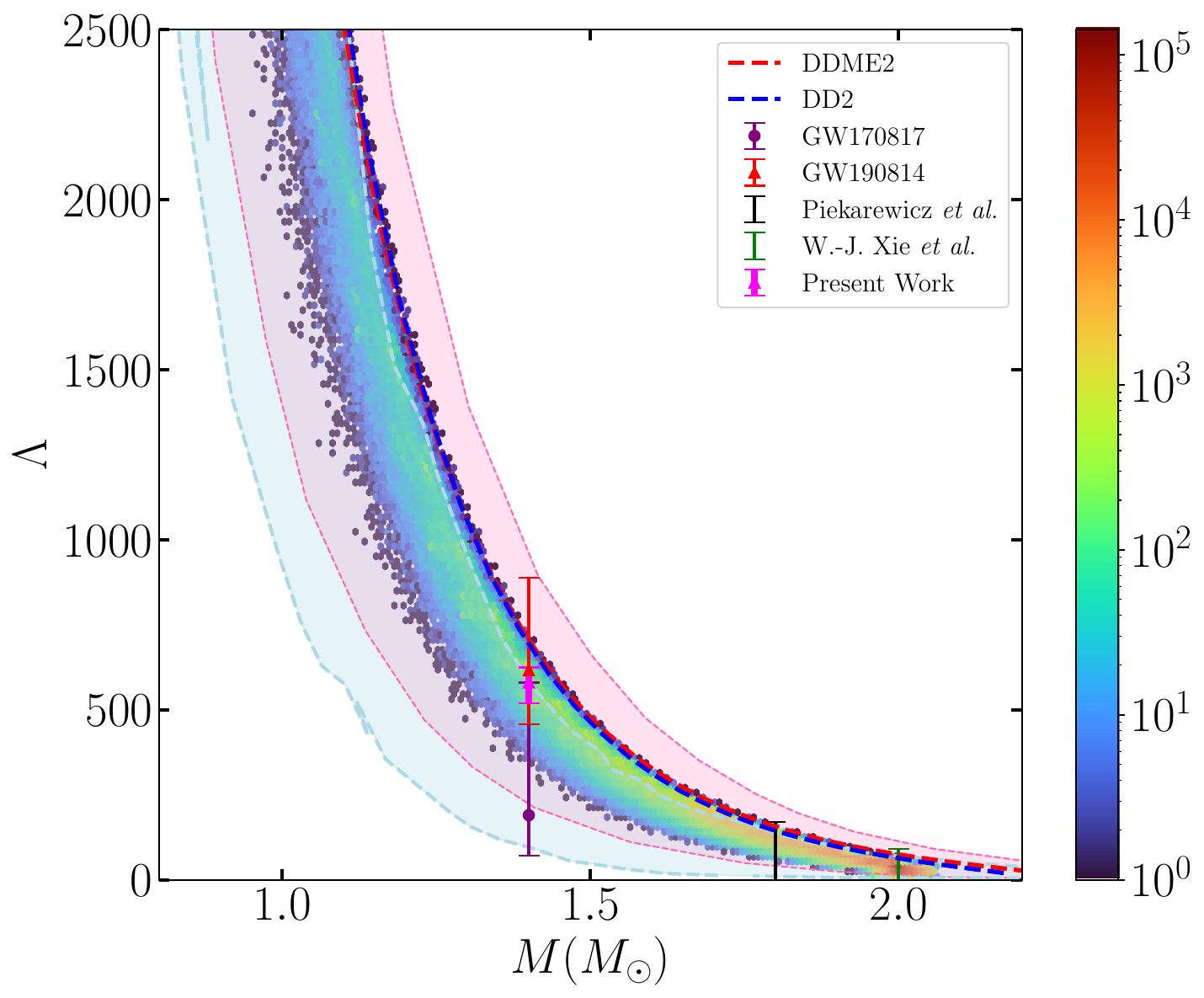}
    \includegraphics[scale=0.35]{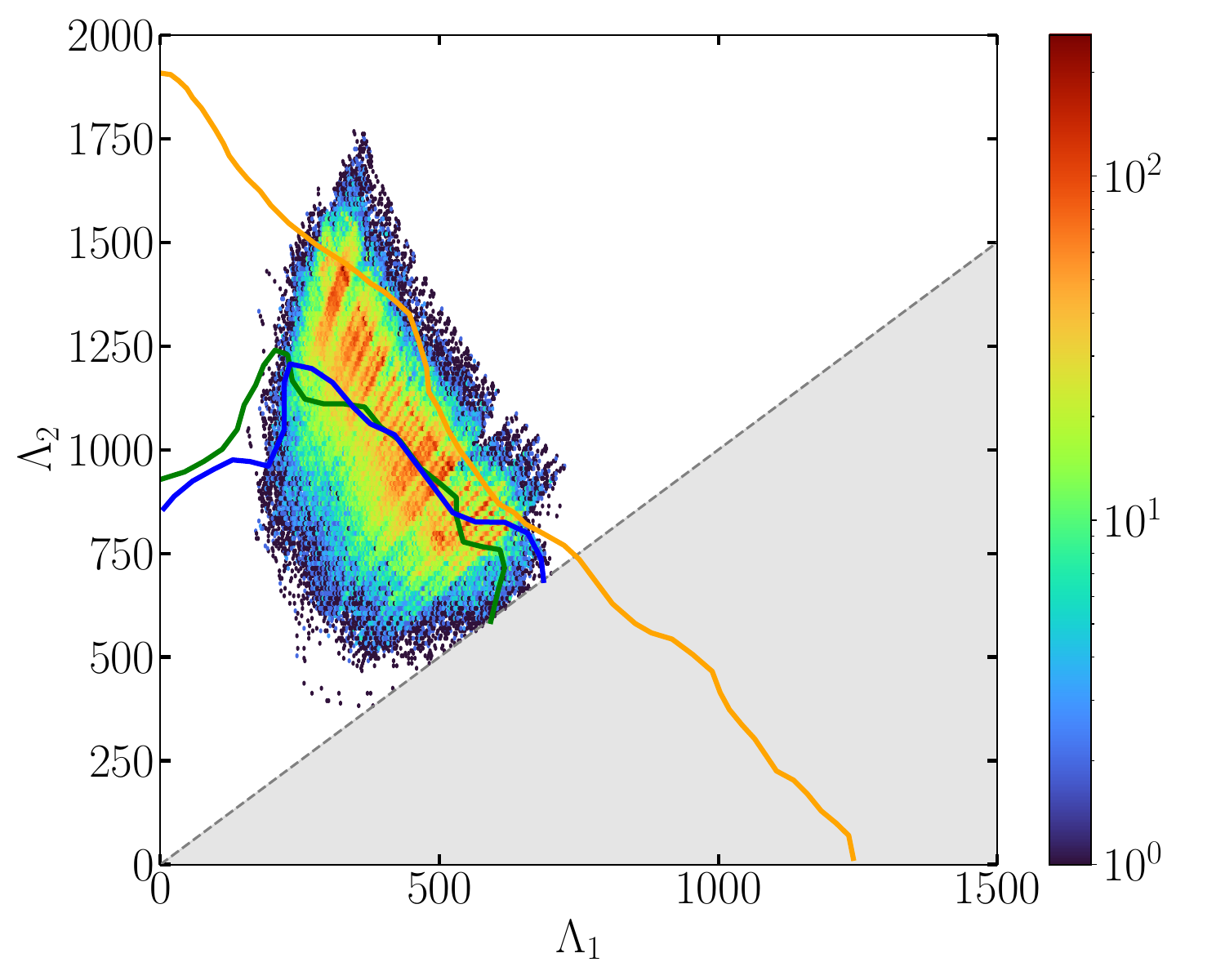}
    \caption{The posterior probability distributions for tidal deformability and mass, denoted as $P(M, \Lambda)$ (left panel) and $P(\Lambda_1, \Lambda_2)$ (right panel), respectively, are illustrated. Here, $\Lambda_1$ and $\Lambda_2$ refer to the dimensionless tidal deformability parameters associated with the binary neutron star merger observed in the GW170817 event. In the left panel, the shaded regions in pink and blue represent the 90\% CI for $\Lambda$ and $M$, respectively, utilizing a parametrized EOS with and without a maximum mass constraint derived from the GW170817 event \cite{Abbott_2018}. Additionally, the purple and red lines depict the observational limits on $\Lambda_{1.4}$ derived from the GW170817 and GW190814 events. Furthermore, black and green error bars denote the values of $\Lambda_{1.8}$ \cite{Piekarewicz_2019} and $\Lambda_{2.0}$ \cite{Xie_2023} obtained from several established theoretical models. The red and blue dashed line corresponds to the EOS using the model DDME2 and DD2 at the median $U_{\bar{K}}$. The right panel showcases the unphysical region where $\Lambda_1$ is less than $\Lambda_2$, illustrated in gray shading. Meanwhile, the green, blue, and orange lines signify the 50\% (dashed) and 90\% (solid) credible levels for the posterior distributions obtained using EOS-insensitive relations, a parametrized EOS without a maximum mass requirement, and independent EOSs, respectively, from the GW170817 event \cite{Abbott_2018}. In both the panel, the colorbar represents the probability on log scale. }
    \label{fig:tidal}
\end{figure*}

\begin{figure*}
    \centering
    \includegraphics[scale=0.7]{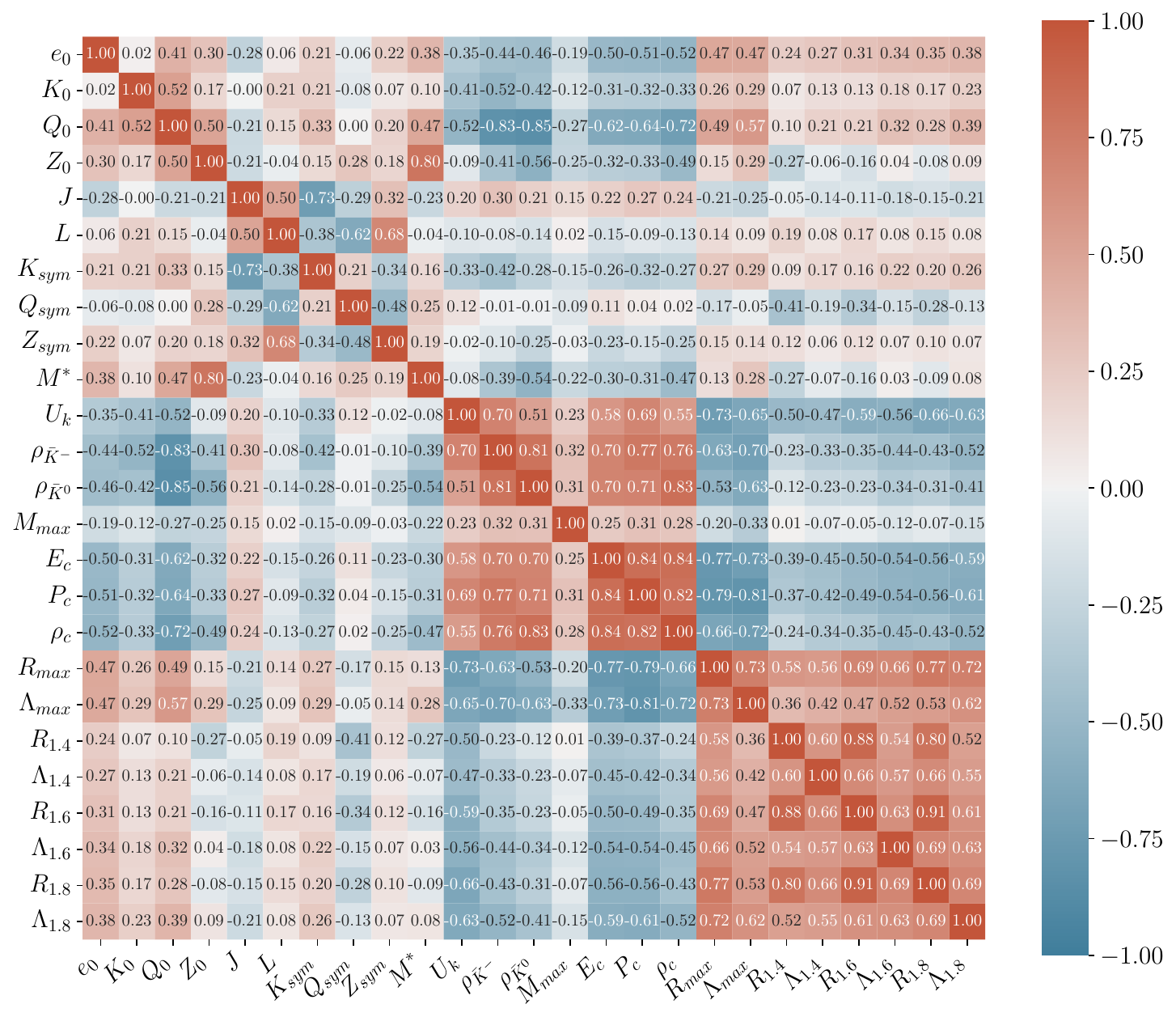}

    \caption{Kendall rank correlation matrix among various nuclear matter properties at the saturation density (i.e. binding energy ($e_0$), incompressibility coefficient $K_0$, the skewness $Q_0$, and the kurtosis $Z_0$, symmetry energy coefficient $J$, the slope $L$, the curvature $K_{sym}$, the skewness $Q_{sym}$, (and the kurtosis $Z_{sym}$) and effective mass ($M^*$),  antikaon potential $(U_{\bar{K}})$, onset density for $K^-$ $(\rho^c_{K^-})$, onset density for $\bar{K}^0$ $(\rho^c_{\bar{K}^0})$ and the neutron star properties with  maximum mass ($M_{max}$), central energy, pressure and density $E_c$, $P_c$ and  $\rho_c$ respectively, radius corresponding to maximum mass ($R_{max}$), tidal defomability corresponding to maximum mass ($\Lambda_{max}$), and the radius and tidal deformability for neutron star mass 1.6 and 1.8 M$_\odot$. The color bar represents the strength of the correlation, with blue representing a strong negative and orange representing a strong positive correlation. The value in white text represent strong correlations.}
    \label{fig:corr}
\end{figure*}

In Fig. \ref{fig:tidal}, we show the posterior probability distributions for tidal deformability and mass, denoted as $P(M, \Lambda)$ (left panel) and $P(\Lambda_1, \Lambda_2)$ (right panel), respectively.  For the tidal deformability ($\Lambda$) and mass ($M$) posterior, we compare the results with the 90\%  CI for $\Lambda$ and $M$, utilizing a parametrized EOS with (pink region) and without (blue region) a maximum mass constraint as derived from the GW170817 event\footnote{\url{https://dcc.ligo.org/LIGO-P1800115/public}} \cite{Abbott_2018}. We also show the constraint imposed in Abbott \textit{et al.} \cite{Abbott_2018} for $\Lambda_{1.4}$, along with the $\Lambda_{1.8}$ \cite{Piekarewicz_2019} and  $\Lambda_{2.0}$ \cite{Xie_2023} constraints determine from  various   established theoretical models which simultaneously satisfy the  data on finite nuclei as well as $2 M_\odot$ constraint. For reference, we also show the $\Lambda_{1.4}$ obtained from the GW190814 event \cite{Abbott_2020}, although we have not used it as a constraint in our Bayesian analysis. 

Our posterior distribution of $\Lambda$ vs $M$ shows a reasonable agreement with the 90\% confidence interval obtained from the GW170817 event. While it falls within the range obtained using a parametrized EOS with a maximum mass constraint of $M \ge 1.97M_\odot$, it leans toward the higher end compared to the range obtained without such a constraint. Specifically, our value for the dimensionless tidal deformability parameter $\Lambda_{1.4}$, considering the antikaon condensation within neutron star interiors, is estimated as $580.098^{+44.35}_{-61.38}$. This value is in reasonably good alignment with the constraints set forth by \cite{Abbott_2018} for the GW170817 event, although its upper limit slightly exceeds the constraints of the aforementioned study. Moreover, our calculated $\Lambda_{1.4}$ also meets the constraints imposed by the GW190814 event \cite{Abbott_2020} and satisfies the well-known constraint $\Lambda_{1.4} < 800$ as reported in \cite{PhysRevLett.119.161101}. Additionally, our posterior distribution aligns with the available constraints on $\Lambda_{1.8}$  \cite{Piekarewicz_2019}  and $\Lambda_{2.0}$ \cite{Xie_2023}. It's worth noting that our calculated $\Lambda_{1.4}$ in this work is relatively higher compared to EOS models with only hadronic degrees of freedom \cite{Malik_2022_1, Mikhail_2023}. This discrepancy can be attributed to the relatively stiffer EOS preferred in the density range crucial for 1.4$M_\odot$ stars in our analysis. Similar to Fig. \ref{fig:eos} and \ref{fig:mr}, we also compare the tidal deformability calculated using the DDME2 and DD2 models. It can be seen that neither of these models satisfies the $\Lambda_{1.4}$ constraints from the GW170817 event, and both lie at the extremes of our posterior. Like the M-R profile, these two sets also do not fall within the posterior calculated in \cite{Malik_2022_1} for the $M$ vs $\Lambda$ relationship. 

In the right panel of Fig. \ref{fig:tidal}, we show the probability distributions  $P(\Lambda_1, \Lambda_2)$, where, $\Lambda_1$ and $\Lambda_2$ refer to the dimensionless tidal deformability parameters associated with the binary neutron star (BNS) merger observed in the GW170817 event. We use the chrip mass $\mathcal{M}=1.188$ and the mass ratio $q=m2/m1$ $(0.7<q<1)$ to calculate the probability distribution  $P(\Lambda_1, \Lambda_2)$. We fix the $m_1$  and calculate the $m_2$ such that $\mathcal{M}=1.188$, $(0.7<q<1)$ and $ 2.73 \le  m_1+m_2 \le 2.78$ and then calculate the value of $\Lambda_1$,\ and  $\Lambda_2$ for the calculated $m_1$ and $m_2$. The result is compared with  posterior distributions obtained using EOS-insensitive relations, a parametrized EOS without a maximum mass requirement, and independent EOSs from the GW170817 event \cite{Abbott_2018}. 
It is seen that $P(\Lambda_1, \Lambda_2)$ obtained by considering the antikaon condensation in the NS interior is in  agreement with the GW170817 results. The maximum probability region lies well inside the 90\% CI obtained from the GW170817 event. Recently, it has been demonstrated that the signature of the Quark-Hadron Phase Transition can be observed in the plot of the two tidal deformabilities $\Lambda_1$ and $\Lambda_2$ of binary neutron stars \cite{Christian_2019}, as well as in the weighted average tidal deformability $\bar{\Lambda}$ at a given chirp mass $\mathcal{M}$ \cite{Sophia_2019}, and in General-Relativistic Neutron-Star Mergers \cite{Elias_2019}. However, for such a signature to be visible, the transition should occur at masses lower than the upper limit on the primary mass, i.e., $m_1 =$ [1.36, 1.60] \cite{Sophia_2019}. 
  \textcolor{black}{Furthermore, it is shown that if a phase transition occurs at several times the nuclear saturation density after the merger, the resulting post-merger signal, which significantly deviates from the inspiral phase that probes the hadronic part of the equations of state, serves as a clear signature of the hadron-quark phase transition. This transition could also cause an earlier collapse of the merged object. \cite{Elias_2019, Fujimoto_2023}}. Similarly, in the current scenario, the antikaon condensation occurs only for masses greater than 1.6 M$_\odot$ (see Fig. \ref{fig:eos}). Therefore, the signatures of antikaon condensation are not visible, and additional pairs of $\Lambda_1$ and $\Lambda_2$ in future gravitational wave detections may help identify any such signatures. 
However, the behaviour of NS tidal deformabilities in the presence of antikaon condenstaion can be theoretically studied for more insight.  Furthermore, we have also examined our posterior for the probability distribution of the weighted average tidal deformability $\bar{\Lambda}$ and the $q$ value, and it complies with the available constraints on $\bar{\Lambda}$ ($\bar{\Lambda}(\mathcal{M}=1.18) =300^{+500}_{-190}$ \cite{PhysRevX.9.011001}).

We now discuss the general correlations among the posterior distributions of various outputs of our Bayesian analysis considering the antikaon condensation in the interior of a neutron star. In Fig. \ref{fig:corr}, we show the Kendall rank correlation matrix among various nuclear matter properties at the saturation density (i.e. binding energy ($e_0$), incompressibility coefficient $K_0$, the skewness $Q_0$, and the kurtosis $Z_0$, symmetry energy coefficient $J$, the slope $L$, the curvature $K_{sym}$, the skewness $Q_{sym}$, (and the kurtosis $Z_{sym}$) and effective mass ($M^*$),  antikaon potential $(U_{\bar{K}})$, onset density for $K^-$ $(\rho^c_{K^-})$, onset density for $\bar{K}^0$ $(\rho^c_{\bar{K}^0})$ and the neutron star properties with  maximum mass ($M_{max}$), central energy, pressure and density $E_c$, $P_c$ and  $\rho_c$ respectively, radius corresponding to maximum mass ($R_{max}$), tidal defomability corresponding to maximum mass ($\Lambda_{max}$), and the radius and tidal deformability for neutron star mass 1.6 and 1.8 M$_\odot$. We use the Kendall rank correlation because the nature of the relationship can differ from the linear, which is the requirement of commonly used Pearson coefficients \cite{KENDALL_1938}. 

In the literature, several studies have explored correlations among nuclear matter properties at the saturation density, astrophysical observables, or their linear combinations \cite{Alam_2016, Margueron_2018, Malik_2018, WEISSENBORN201262, Hornick_2018, Tsang_2012, Mikhail_2023, PhysRevD.107.103054} in search of universal correlations or those dependent on setup or model. Within nuclear matter properties at the saturation density, notable correlations include $M^*-Q_0$, $M^*-Z_0$, $J-K_{sym}$, $L-Z_{sym}$, and $L-Q_{sym}$ pairs. Notably, a strong positive correlation exists between $M^*-Z_0$ and a negative correlation between $L-Q_{sym}$, as also observed in Ref. \cite{Mikhail_2023}, which explores the dense matter EOS using the same formalism and Bayesian approach as this study. Furthermore, strong correlations between $L-Z_{sym}$ (positive) and $L-Q_{sym}$ (negative) are evident from the corner plot in \cite{Malik_2022_1}, indicating their universal nature even in hyperonic matter, as shown in \cite{Malik_2022}. However, weak correlation between $L-K_{sym}$ contrasts with findings in \cite{Tews_2017, 2009PhRvC..80d5806V}, as observed by us and also reported in Refs. \cite{Mikhail_2023, Malik_2022, Malik_2022_1}. We also found no correlation between isoscalar and isovector parameters. 

Regarding the antikaon potential $(U_{\bar{K}})$ and onset densities for $K^-$ $(\rho^c_{K^-})$ and $\bar{K}^0$ $(\rho^c_{\bar{K}^0})$, these exhibit weak correlations with isovector properties ($J$ and higher-order derivatives) but relatively strong negative correlations (confirmed with Pearson correlation values greater than 0.7) with isoscalar ($K_0$) and higher-order derivatives. 
$U_{\bar{K}}$ also shows strong correlation with astrophysical observables except for the canonical case, likely due to the absence of antikaon condensation in such stars. The strength of correlation increases notably for heavier NSs, highlighting the significance of antikaons in such scenarios.

We observe a strong negative correlation between central density ($\rho_c$), pressure ($P_m$), and the isoscalar parameter $Q_0$. Using Pearson correlation statistics, we also identify a strong negative correlation with $K_0$, consistent with findings in \cite{Mikhail_2023} (refer to Fig. 8 of \cite{Mikhail_2023}). Additionally, $\rho_c$ and $P_c$ show a strong positive correlation with the strength of the antikaon potential and the onset density of $K^-$ $(\rho^c_{K^-})$ and $\bar{K}^0$ $(\rho^c_{\bar{K}^0})$. Notably, this correlation is strongest for $(\rho^c_{\bar{K}^0})$, which is intuitive since an early onset of $\bar{K}^0$ softens the EOS compared to its late appearance. However, we find no significant correlation between maximum mass and any parameter in Fig. \ref{fig:corr}. Similarly, in our study, we do not observe significant correlations for $R_{1.4}$ and $\Lambda_{1.4}$ with any nuclear matter parameter. This lack of correlation contrasts with reported correlations in the literature involving $R_{1.4}$ and $\Lambda_{1.4}$ \cite{malik_2020, Pradhan_2022, Tsang_2012, Tsang_2020, Carson_2019}, which heavily depend on the distribution of nuclear matter parameters and are essentially model or setup dependent \cite{Patra_2023, Kunjipurayil_2022}.

\section{\label{summary} Summary}
In this study, we employ a DDRH field model within a Bayesian framework to investigate the EOS of dense matter, with a specific focus on antikaon condensation involving $K^-$ and $\bar{K}^0$ within NSs.
\textcolor{black}{Due to the huge uncertainties and absence of reliable experimental data of kaon-nucleon scattering, it is quite difficult to ascertain the presence as well as influence of antikaons in dense matter.}
Our primary objective is to pin down the antikaon potential range arising from phenomenological estimations and to quantify the same in a manner that aligns with the constraints derived from $\chi$EFT calculations, nuclear saturation properties, and astrophysical observations from pulsars PSR J0030+0451, PSR J0740+66, and the GW170817 event. Additionally, we analyze the impact of various constraints from nuclear physics and astrophysics on the EOS of dense matter. We also investigate correlations among input parameters of the model, parameters of nuclear matter, and selected key properties of NSs.

We utilize a modified version of DDRH, as discussed in \cite{Mikhail_2023, Malik_2022}, featuring fewer input parameters compared to conventional models like DDME2 and DD2 for the inclusion of antikaons in the DDRH model, the meson-antikaon couplings are assumed to be non-density-dependent, while the vector coupling parameters in the kaon sector are determined using the iso-spin counting rule and quark model \cite{2001PhRvC..64e5805B}. 
Previous studies, such as Ref. \cite{Kheto2023}, have estimated the antikaon potential to be around -130 MeV and examined its effects on the stability and oscillation modes of NSs.
\textcolor{black}{Another study \cite{10.1093/mnras/stx2999} investigated the impact of antikaons on superfluid matter and suggested the optical potential range to be $-80$ to $-130$ MeV. Ref.-\cite{Vivek_2020} also suggested the antikaon potential to be deeper than $-120$ MeV.}
Our Bayesian analysis establishes the antikaon potential within 68(90)\% confidence intervals as $-129.36^{+12.53(+32.617)}_{-3.837(-5.696)}$ MeV. 
The posterior distribution from our Bayesian analysis aligns with recent experimental and theoretical constraints on nuclear matter properties at the saturation density. 
NS properties, such as maximum mass, radii, and tidal deformabilities, are consistent with recent observational data. While antikaon condensation is not favored within canonical NSs, those with masses exceeding 2 M$_\odot$ seem to support the possibility of both $K^-$ and $\bar{K}^0$ condensation. 
However, we will require more data on NS mass and tidal deformability from future advanced telescopes and detectors, especially in the range $M=[1.4-2]M_\odot$, in order to better constrain the EOS and search for signatures of antikaon condensation if present. Our calculations indicate the onset density of $K^-$ and $\bar{K}^0$ to be within 68(90)\% CI at $2.80^{+0.35(+1.00)}_{-0.05(-0.10)}$ and $4.45^{+0.70(2.65)}_{-0.20(-0.25)}$ in units of $\rho_0$.

Upon the onset of $K^-$ and $\bar{K}^0$ condensation, the EOS undergoes a noticeable softening, particularly beyond the onset density of $\bar{K}^0$, contrasting with other EOS configurations such as pure hadronic or hyperonic cases. Consequently, the median maximum mass value hovers around $2$ $M_\odot$. However, to meet the lower bound requirement ($M>2M_\odot$), a relatively stiffer EOS is preferable near $\rho_0$, which then softens with the emergence of $K^-$ and $\bar{K}^0$ condensation. This softening significantly impacts the speed of sound within the neutron star's interior, especially assuming antikaon condensation, leading to a notably lower speed of sound compared to models focusing solely on hadronic or hyperonic degrees of freedom. The EOS posterior aligns well with recent Bayesian estimations derived from microscopic and multi-messenger astrophysics, with the mass-radius (M-R) posterior satisfying both GW and NICER posterior estimations. 
The tidal deformability agrees with the GW170817 event, although our estimation of $\Lambda_{1.4}$ in this work is relatively higher compared to EOS models with only hadronic degrees of freedom \cite{Malik_2022_1, Mikhail_2023}. This difference can be attributed to the relatively stiffer EOS preferred in the density range crucial for 1.4$M_\odot$ stars in our analysis. It is seen that our posterior of the EOS is similar to the DDME2 and DD2 models, although they lie outside our posterior for the EOS, M-R profile, and $\Lambda-M$ relationship.

Finally, we analyze the correlations among various nuclear matter antikaon and NS properties. We found no correlation between isoscalar and isovector nuclear matter properties. Among the nuclear matter parameters, the most notable correlations are between pairs such as $M-Q_0$, $M-Z_0$, $J-K_{sym}$, $L-Z_{sym}$, and $L-Q_{sym}$. These correlations are also well-documented in the literature, suggesting their universal nature across different EOSs considering various degrees of freedom. In terms of the antikaon potential $(U_{\bar{K}})$ and the onset densities for $K^-$ $(\rho^c_{K^-})$ and $\bar{K}^0$ $(\rho^c_{\bar{K}^0})$, we observe weak correlations with isovector properties but relatively strong negative correlations with isoscalar properties. 
$U_{\bar{K}}$ also demonstrates a strong correlation with astrophysical observables, except for the canonical case. Additionally, $\rho_c$ and $P_c$ exhibit a strong positive correlation with the strength of the antikaon potential and the onset density of $K^-$ $(\rho^c_{K^-})$ and $\bar{K}^0$ $(\rho^c_{\bar{K}^0})$.

In this study, we have considered the EOS to be composed of nucleons ($n$ and $p$) along with antikaons ($K^-$ and $\bar{K}^0$). Hyperons or massive $\Delta$ resonances were not included due to the fact that the presence of these species significantly delays the onset of condensation \cite{Thapa_2021}. Additionally, $\bar{K}^0$ condensation appears only for very deep antikaon potentials and at densities $5-6$ times the nuclear saturation densities. 
In presence of the exotic particle spectrum such as hyperons, $\Delta-$baryons, the onset of antikaon condensation will be shifted to higher matter densities on the basis of their threshold conditions in dense matter.
%suggesting that they may not be present within the central neutron star density if all the species are accounted for. 

\section*{Acknowledgement}
V. P. and M. S. acknowledge the financial support from the Science and
Engineering Research Board, Department of Science andTechnology, Government of India through Project
No. CRG/2022/000069 and the HPC cluster facility at IIT Jodhpur. V.P. is also thankful to Chun Huang, Ming-Zhe Han and Pratik Thakur for helping in the Bayesian framework. DB acknowledges the hospitality at  Frankfurt Institute for Advanced Studies.

\bibliographystyle{apsrev4-2}
\bibliography{kaon}% Produces the bibliography via BibTeX.

\end{document}